\documentclass[useAMS,usenatbib]{aa}  

\usepackage{txfonts}
\usepackage{graphicx}
\usepackage{amsmath}
\usepackage{tabularx}
\usepackage{inputenc}
\usepackage{supertabular}
\usepackage{ltxtable}
\usepackage{subfig}
\begin{document}

   \title{Magnetic cycles and rotation periods of late-type stars from photometric time series}
	\titlerunning{Magnetic cycles and rotation periods}

   \author{A. Su\'{a}rez Mascare\~{n}o
          \inst{1,2}
          \and
         R. Rebolo
         \inst{1,2,3}
         \and
         J.~I. Gonz\'alez Hern\'andez
         \inst{1,2}
         \fnmsep
          } 

   \institute{Instituto de Astrof\'{i}sica de Canarias, E-38205 La Laguna, Tenerife, Spain\\
              \email{asm@iac.es}
         \and
             Universidad de La Laguna, Dpto. Astrof\'{i}sica, E-38206 La Laguna, Tenerife, Spain
 \and
             Consejo Superior de Investigaciones Cient{\'\i}ficas, E-28006 Madrid, Spain\\
             }

   \date{Written November 2015 to March 2016 -- Revised June-July 2016}

 
  \abstract
   {}
   {We investigate the photometric modulation induced by magnetic activity cycles  and study  the relationship between  rotation period and activity cycle(s) in late-type (FGKM) stars. }
   {We analysed  light curves, spanning up to nine years, of 125 nearby stars  provided by the All Sky Automated Survey (ASAS). The sample is mainly composed of low-activity, main-sequence late-A to mid-M-type stars. We performed a search for short (days) and long-term (years)  periodic variations in the photometry. We modelled  the light curves with combinations of sinusoids to measure   the properties of these periodic signals. To provide a better statistical interpretation of our results, we complement our new results  with  results from previous similar works.}
   {We have been able to measure long-term photometric cycles of 47 stars, out of which 39 have been derived  with false alarm probabilities (FAP) of less than 0.1 per cent. Rotational modulation was also detected and rotational periods were measured  in 36 stars. For 28 stars we have simultaneous measurements of   activity cycles and rotational periods, 17 of which are M-type stars.  We measured both  photometric amplitudes and periods from sinusoidal fits. The measured cycle periods range from 2 to 14 yr with photometric amplitudes in the range of 5-20 mmag. We found that the distribution of cycle lengths for the different spectral types is similar, as the mean cycle is 9.5 years for F-type stars, 6.7 years for G-type stars, 8.5 years for K-type stars, 6.0 years for early M-type stars, and 7.1 years for mid-M-type stars.  On the other hand, the distribution of rotation periods is completely different, trending to longer periods for later type stars, from a mean rotation of 8.6 days for F-type stars to 85.4 days in mid-M-type stars. The amplitudes induced by magnetic cycles and rotation show a clear correlation. A trend of photometric amplitudes with rotation period is also outlined in the data.  The  amplitudes of the photometric variability induced by activity cycles  of main-sequence  GK stars are lower than those of early- and mid-M dwarfs   for a given activity index. Using spectroscopic data, we also provide an update in the empirical relationship between the level of chromospheric activity as given by $\log_{10}R'_{HK}$  and the rotation periods.}
   {}


   \maketitle
%
\section{Introduction}

It is widely recognised that starspots in late-type dwarf stars  lead to periodic light variations associated with the rotation of the stars \citet{Kron1947}. Starspots trace magnetic flux tube emergence and provide valuable information on the forces acting on flux tubes and photospheric motions, both important agents in the dynamo theory (Parker 1955, Steenbeck et al. 1966, Bonanno et al. 2002). Rotation plays a crucial role in the generation of stellar activity \citep{Skumanich1972}. This becomes evident from the strong correlation of magnetic activity indicators with rotation periods \citep{Pallavicini1981, WalterBowyer1981, Vaughan1981, Middelkoop1981, Mekkaden1985, Vilhu1984, SimonFekel1987, Drake1989,Montes2004,Dumusque2011, Dumusque2012,Masca2015}. 
Stellar rotation coupled with convective motions generate strong magnetic fields in the stellar interior and produce different magnetic phenomena, including starspots in the photosphere. Big spotted areas consist of groups of small spots whose lifetime is not always easy to estimate, but the main structure can survive for many rotations, causing the coherent brightness variations that we can measure \citep{HallHenry1994}. In solar-like main-sequence  stars the light modulations associated with rotation  are of order a few percent (Dorren and Guinan 1982, Radick et al. 1983), while  in young fast rotating stars these modulations can be significantly larger (e.g. Frasca et al. 2011). Starspot induced light modulation was also proposed for M dwarfs decades ago \citep{Chugainov1971} and more recently  investigated by, for example \citet{Irwin2011}, \citet{Kiraga2012} and \citet{West2015}. 

Long-term  magnetic activity similar to that of the Sun is also observed on stars with external convection envelopes \citep{Baliunas1985, Radick1990, Baliunas1996,Strassmeier1997,Lovis2011,Savanov2012,Robertson2013}. Photometric and spectroscopic time series observations over decades have revealed stellar cycles similar to the 11 yr sunspot cycle. In some active stars even multiple cycles are often observed \citep{BerdyuginaTuominen1998,BerdyuginaJarvinen2005}. It is possible to distinguish between cycles that are responsible for overall oscillation of the global level of the activity (similar to the 11 yr solar cycle) and cycles that are responsible for the spatial rearrangement of the active regions (flip-flop cycles) at a given activity level, such as the 3.7 yr cycle in sunspots \citep{BerdyuginaUsoskin2003, Moss2004}. The correct understanding of the different types of stellar variability, their relationships, and their link to stellar parameters is a key aspect to properly understand the behaviour of the stellar dynamo and its dependence on stellar mass. While extensive work has been conducted in FGK stars over many decades,  M dwarfs  have not received so much attention with only a few tens  of long-term activity cycles reported in the literature \citep{Robertson2013, GomesdaSilva2012}.

Understanding the full frame of stellar variability in stars with a  low level of activity  is also crucial for exoplanet surveys. Modern spectrographs can now reach sub ms$^{-1}$ precision in the radial velocity measurements \citep{Pepe2011} and next generation instrumentation  is expected to reach a precision of a few cms$^{-1}$ \citep{Pepe2013}. At such precision level,  the stellar activity induced signals in the radial velocity curves  become a very important  limiting factor in the search for Earth-like planets. Activity induced signals  in timescales of days associated with rotation and in  timescales of years for magnetic cycles are two of the most prominent sources of radial velocity induced signals \citep{Dumusque2011}. Measuring and understanding this short-term and long-term variability in different types of stars  and the associated effects in radial velocity curves, and being able to predict the behaviour of a star, is required to  disentangle stellar induced signals from Keplerian signals.  

 Ground-based automatic photometric telescopes have been running for decades, providing the photometric  precision and time coverage to explore rotation periods and  activity cycles for  sufficiently bright stars with a low activity level. In this work, using  All Sky Automated Survey (ASAS)  available  long-term series of photometric data,  we attempt the determination of rotational periods and long-term activity cycles of a  sample of G to mid-M stars, with emphasis on the less studied  low activity  M dwarfs. We investigate the relationships between the level of magnetic  activity derived from Ca II H\&K, and  the rotation and activity cycle periods derived from the photometric series. Several studies have  taken advantage of the excellent  quality of the Kepler space telescope light curves to determine  activity cycles and rotation period, for example \citet{Vida2014}, however  the limited time span of the observations restricts  these studies to   cycles of only a few years. Here we investigate the presence of significantly longer activity cycles in our sample of stars.


\section{Sample and data}

Our sample consists of the stars in the study of  rotational modulation of the Ca II lines by \citet{Masca2015} supplemented with a selection of nearby (suitably bright) mostly  southern stars with reliable  long photometric series  in  the ASAS public database. We selected  the brightest examples of stars in the ASAS database, which offered photometric data in a linear regime. A  few young very active fast rotating stars were also included  to act as anchor points in the fast-rotating end, high activity range. In total  we have analysed photometric data from  $125$ stars with spectral types  from late F to mid-M type with a goal to determine the photometric modulation induced by rotation and long-term activity cycles. Most of the stars in the sample  are main-sequence stars.   

\begin{figure}
\includegraphics[width=9.0cm]{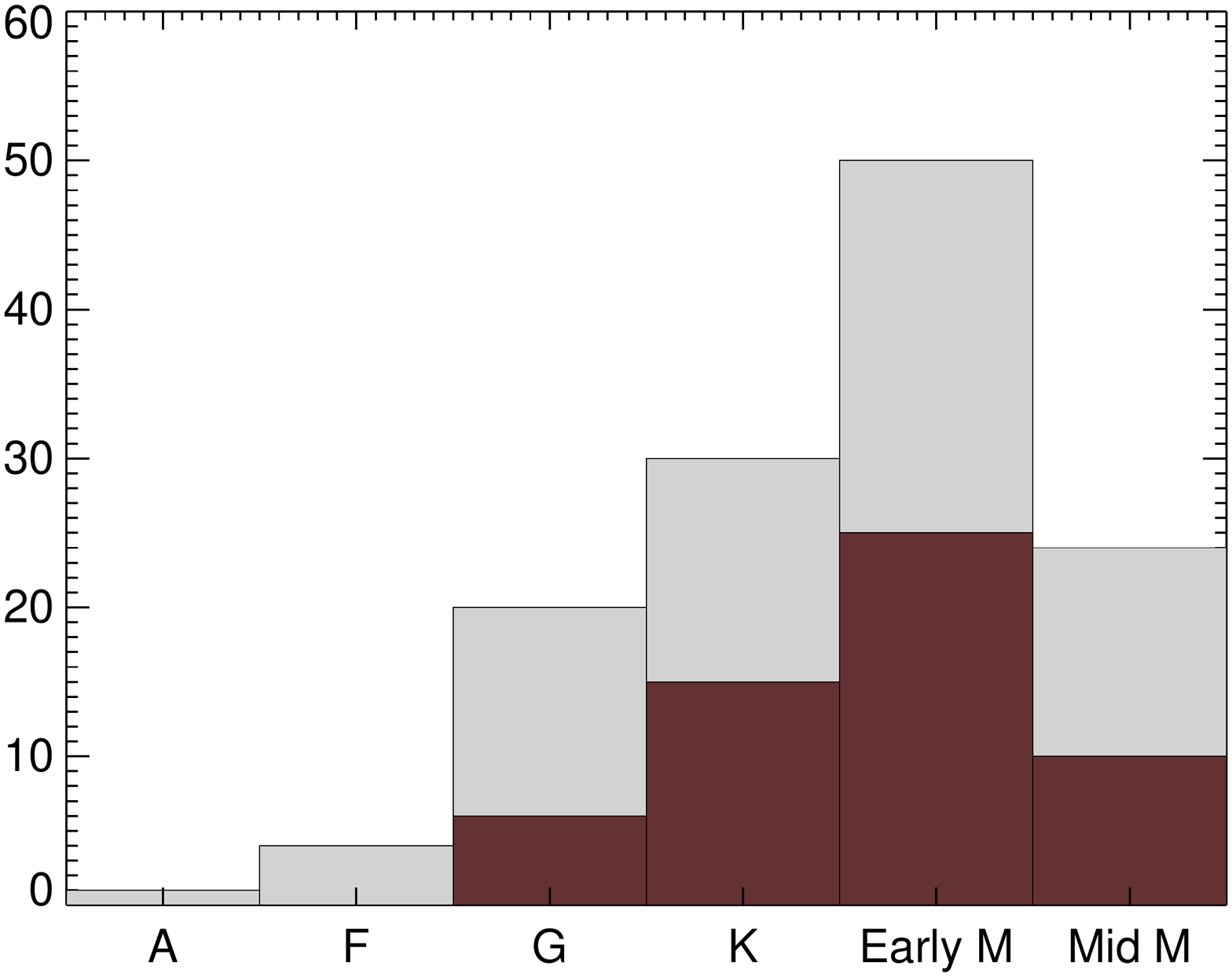}
\caption{Distribution of spectral types in our sample. M dwarfs were separated in two groups with    the fully convective dwarfs (spectral type M3.5 and later) in a separated bin. Filled bins indicate  stars in which we were able to retrieve long-term photometric trends and cycles  likely caused by stellar activity.}
\label{sp_dis}
\end{figure}

Figure ~\ref{sp_dis} shows the distribution of spectral types for the total sample of stars originally considered. Table ~\ref{data_sample} lists relevant data for only those  stars where a reliable determination of photometric modulation  was achieved (see below). 

\subsection{Photometric Data}

The all sky survey ASAS \citep{Pojmanski1997}  in the $V$ and $I$ bands has been running at Las Campanas Observatory, Chile since 1998. It has a spatial scale  of $14"/pixel$ and an average accuracy of $\sim0.05$ mag per exposure. Best photometric results are achieved for stars with V $\sim$8-12, but this range can be extended if some additional  control on the quality of the data is implemented. ASAS has produced light curves for around $10^{7}$ stars at $DEC < 28^{\circ}$. The catalogue supplies ready-to-use light curves with flags indicating the quality of the data. For this analysis we relied  only on  good quality data (grade "A" and "B" in the internal flags). Even after this quality control,  there are still some high dispersion measurements  that cannot  be explained by a "regular" stellar behaviour. There are  some cases of extreme scatter in stars closer to the magnitude  limits of the survey and some stars that show a behaviour compatible with stellar flares. As our data is not well suited for modelling flares,  to remove  flare affected points,  we iteratively rejected all measurements that deviate from the median value more than $2.5$ times the rms of the full  light curve. Iteration was applied  until no more measurements where left  outside  these limits.

\subsection{Spectroscopic Data}

As proxy for the  chromospheric activity level of the stars in our sample we  rely on  the standard chromospheric activiy index  $\log_{10}R'_{HK}$ \citep{Noyes1984}. This index  was directly adopted  from \citet{Lovis2011} and \citet{Masca2015} for the stars in common with these works. For the remaining stars we searched for  available spectra of the Ca II  H\& K lines  in the HARPS ESO public data archive and performed our own measurement of the index. HARPS \citep{Mayor2003} is a fibre-fed high-resolution spectrograph installed at the 3.6 m ESO telescope in La Silla Observatory (Chile). The instrument has a resolving power $R \sim 115000$ over a spectral range from 378 to 691 nm. The HARPS pipeline provides extracted and wavelength-calibrated spectra, as well as RV measurements. We used the extracted order-by-order wavelength-calibrated spectra produced by the HARPS pipeline for our analysis. In order to minimise the effects related to atmospheric changes, we created a spectral template for each star by deblazing and co-adding every available spectrum and used the co-added spectrum to correct the order-by-order flux ratios for the individual spectra. We  corrected each spectrum for the barycentric radial velocity of the Earth and the radial velocity of the star using the measurements given by the standard pipeline. We finally corrected each spectrum from the flux dispersion and re-binned the spectra into a wavelength-constant step. 

For the measurement of the $\log_{10}R'_{HK}$, we followed \citet{Masca2015}, where an extension of  the index to M-type dwarf stars, which are the majority of our present sample, was proposed. 

We have 
\begin{equation}
   R'_\textrm{HK}=1.34 \cdot 10^{-4} \cdot C_{\rm cf}(B-V) \cdot S-R_{\rm phot}(B-V).
\end{equation}

Where $S$ is flux in the Ca II H\&K lines, defined as
\begin{equation}
   S=\alpha {{N_{H}+N_{K}}\over{N_{R}+N_{V}}},
\end{equation}

where $N_{H}$ and $N_{K}$ are the fluxes in the core of the lines, $N_{R}$ and $N_{V}$ the fluxes on the nearby continuum, and $\alpha$ a constant equal to 2.3 times 8.  

Then we have $C_{CF}$, the bolometric correction, measured as

\begin{equation}
\begin{split}
  C_{\rm cf}=(S_{R}+S_{V}) \cdot 10^{-4} \cdot 10^{04 (m_{v} + BC)},
\end{split}
\end{equation}

where $S_{R}$ and $S_{V}$ are the mean fluxes in the reference continuum and $BC$ the bolometric correction. This quantity can also be estimated as
\begin{equation}
\begin{split}
  \log_{10}C_{\rm cf}=0.668-1.270(B-V)+0.645 (B-V)^{2}\\-0.443(B-V)^{3},
\end{split}
\end{equation}

and $R_{\rm phot}$ is the photospheric contribution to the mean activity level, which we estimate as 
\begin{equation}
\begin{split}
   R_{\rm phot}= 1.48 \cdot 10^{-4} \cdot exp[-4.3658 \cdot (B-V)].
\end{split}
\end{equation}

\begin {table*}
\begin{center}
\caption {Relevant data for stars in our sample with periodic photometric modulation\label{tab:data_sample}}
    \begin{tabular}{ l  l  l l l l l l l l l } \hline
Star &Sp. Type  &  $\log_{10}R'_{HK}$ & $m_{B}$ & $m_{V}$ &$<m_{V_{ASAS}}>$ & $N_{meas}$ & Time Span&  Ref\\ 
        &        &  &  & &&  & (Years) & \\\hline
HD1388          & G0  &  --4.89 $\pm$ 0.01 & 7.09 & 6.50 & 6.50 & 335 & 7.4 & 1, 5,\\
HD10180                 & G1  & --5.00 $\pm$ 0.03 & 7.95 & 7.32 & 7.32 & 524 & 8.8 & 2, 5\\
HD21019         & G2  &  --5.12 $\pm$ 0.02 & 6.90 & 6.20 & 6.20 & 331 & 8.1 & 2, 4\\
HD2071                  & G2  &  --4.89 $\pm$ 0.03 & 7.95& 7.27&  7.27 & 641 & 8.4 & 1, 6 \\
HD1320          & G2  &  --4.79 $\pm$ 0.03 & 8.63 & 7.98 &  7.95 & 614 & 9.0 &  1,3 \\
HD63765                 & G9  &  --4.77 $\pm$ 0.04 & 8.85& 8.10 & 8.09 & 588 & 9.0 & 1, 5\\
HD82558                 & K0    &  --4.01 $\pm$ 0.05 & 8.76 & 7.89 & 7.87 & 481 & 9.0 & 0, 7, 11\\
HD155885                & K1    &  --4.55 $\pm$ 0.02 & 5.96 & 5.08 & 4.96 & 562 & 8.6 & 3, 10\\
HD224789        & K1 &  --4.55 $\pm$ 0.03 & 9.12 & 8.24 & 8.23 & 489 & 9.0 & 1, 5\\
V660 Tau                & K2                    &  & 13.66 & 12.60 & 12.63 & 275 & 7.0 & 18\\
HD176986        & K2.5 &  --4.90 $\pm$ 0.04 & 9.39 & 8.45 & 8.46 & 405 & 8.7 & 1, 5\\
HD32147                 & K3 &  --4.92 $\pm$ 0.04 & 7.27  &  6.21 & 6.19 & 459 & 8.7 & 0, 9\\
HD45088                 & K3    &  --4.48 $\pm$ 0.17 & 7.74 & 6.77 & 6.78 & 437 & 6.9 & 0, 7\\
HD104067        & K3 &  --4.77 $\pm$ 0.04 & 8.90 & 7.92 & 7.91& 795 & 9.0 & 1, 7\\
HD215152        & K3 &  --4.92 $\pm$ 0.06 & 9.13 & 8.13 & 8.09 & 327 & 9.0 & 1, 5\\
V410 Tau        & K3  &&  12.13 & 10.75 & 10.85 & 56 & 1.1 & 5\\
HD131977        & K4 &  --4.33 $\pm$ 0.13 & 6.83 & 5.72 & 5.66& 228 & 8.3 & 0 9\\
HD125595        & K4 &  --4.82 $\pm$ 0.07 & 10.13 & 9.03& 8.99 & 703 & 8.7 & 1, 6\\
HD118100                & K5     & & 10.55 & 9.37 & 9.37 & 476 & 8.6 & 7\\
GJ9482          & K7  & --4.74 $\pm$ 0.05 & 11.80& 10.38& 10.38& 524 & 8.8 & 0, 7 \\
HD113538        & K9 &  --4.82 $\pm$ 0.07 & 10.43 & 9.06 & 9.05& 994 & 8.8 & 0, 7 \\
GJ846           & M0 & --4.84 $\pm$ 0.04 & 10.62 &  9.15 &  9.15 & 312 & 9.0 & 1, 7\\
GJ676A          & M0 &  --4.96 $\pm$ 0.04 & 11.03 & 9.59 & 9.60 & 789 & 8.7 & 1, 7\\
GJ229           & M1 &  --4.91 $\pm$ 0.04 & 9.61 & 8.13 & 8.15 & 708 & 9.0 & 0, 7\\
HD197481                & M1 &  --4.05 $\pm$ 0.04 & 10.05 & 8.63 & 8.63 & 455 & 8.9 & 0, 4, 11\\
GJ514           & M1 &  --5.10 $\pm$ 0.06 & 10.52 & 9.03 & 9.04 & 341 & 8.0 & 1,7\\
GJ239           & M1            &  & 11.10 & 9.59 & 9.60 & 528 & 7.0 &  7\\
GJ205           & M1.5 &  --4.96 $\pm$ 0.06 & 9.44 & 7.97 & 7.95 & 620 & 9.0 & 1, 7\\
GJ832           & M1.5 & --5.21 $\pm$ 0.07 & 10.18 & 8.67 & 8.67 & 482 & 8.9 & 1, 7\\
GJ536           & M1.5 &  --5.22 $\pm$ 0.07 & 11.18 & 9.71 & 9.71 &345 & 3.6 & 1, 7\\
GJ382           & M2 & --4.80 $\pm$ 0.06 & 10.76 & 9.26 & 9.25 & 494 & 9.0 & 1, 7\\
GJ588           & M2.5 &  --5.34 $\pm$ 0.10 & 10.81 & 9.31 & 9.33 & 503 & 8.7 & 1, 7\\
GJ176           & M2.5 &  --4.99 $\pm$ 0.07 & 11.49 & 9.95 & 10.01 & 413 & 7.0 & 1, 7\\
GJ752A          & M3 &  --5.16 $\pm$ 0.07 & 10.63 & 9.12 & 9.14 & 389 & 8.1 & 1, 7\\
GJ674           & M3 & --5.07 $\pm$ 0.08 & 10.97 & 9.41 & 9.39 & 537 & 8.7 & 1, 7\\
GJ680           & M3 & & 11.67 & 10.13 & 10.13 & 834 & 8.7 & 7\\
GJ408           & M3 & & 11.58 & 10.02 & 10.03 & 273 & 6.5 & 7\\
GJ877           & M3 &  --5.33 $\pm$ 0.15 & 11.87 & 10.38 & 10.39 & 1049 & 9.0 & 0 ,7\\
GJ479           & M3                            &                --4.88 $\pm$0.05               & 12.20 & 10.66   & 10.67 & 534 & 8.7 & 0, 7\\
GJ358           & M3 & --4.69 $\pm$ 0.10 & 12.21 & 10.69& 10.72 & 994 & 9.0 & 1, 7\\
GJ273           & M3.5 &  --5.52 $\pm$ 0.07 & 11.44 & 9.87 & 9.98 & 366 & 8.0 & 1, 7 \\
GJ849           & M3.5 & --5.14 $\pm$ 0.04& 11.87 & 10.37 & 10.36 & 424 & 8.9 & 1, 7 \\
GJ729           & M3.5          &                                                       & 12.24 & 10.50 & 10.54  & 620 & 8.6 & 7\\
GJ896A          & M3.5 &  & 12.27 & 10.35&10.07& 257 & 6.9 & 13, 14\\
GJ317           & M3.5 & --4.57 $\pm$ 0.11 & 13.2 & 11.98 & 12.07 & 561 & 9.0 & 0, 4, 16\\
GJ526           & M4 &  --5.31 $\pm$ 0.07 & 9.92 & 8.43 & 8.45& 281 & 6.5 & 1, 7\\
GJ628           & M4 &  --5.47 $\pm$ 0.11 & 11.64 & 10.07 & 10.12 & 498 & 8.7 & 0, 17\\
GJ234           & M4+M5.5 & --4.60 $\pm$ 0.24 & 12.76 & 11.07 & 11.08  & 450 & 8.7 & 0, 7 \\
GJ54.1          & M4            &               & 13.89 & 12.07 & 12.09 & 435 & 9.0 & 7\\
LP 816-60       & M4 & & 13.03 & 11.46 & 11.44 & 413 & 8.5 & 7 \\
GJ285           & M4.5 & --4.09 $\pm$ 0.06 & 12.83 & 11.23 & 11.24 & 530 & 8.7 & 0, 7\\
GJ447           & M4.5 & --5.73 $\pm$ 0.07 & 12.91 & 11.15 & 11.12& 364 & 8.7 & 0, 17\\
GJ581           & M5 &  --5.79 $\pm$ 0.03 & 12.17 & 10.57 & 10.57 & 687 & 8.7 & 1, 7 \\
GJ551           & M5.5 &  --5.65 $\pm$ 0.17 & 12.97 & 11.10 & 11.23  & 991 & 8.7 & 1, 8 \\
GJ406           & M6            &                                                       & 15.54 & 13.51 & 13.51 & 331 & 8.0 &  17\\

 \hline
\label{data_sample}
\end{tabular}  
\end{center}
References for $\log_{10}R'_{HK}$:
0) This work, 
1) \citet{Masca2015},
2) \citet{Lovis2011},
3) \citet{Baliunas1995} \\
References for magnitudes:
4) \citet{Mermilliod1986},
5) \citet{Hog2000},
6 ) \citet{Cousins1962},
7) \citet{Koen2010},
8) \citet{JaoHenry2014},
9) \citet{Ducati2002},
10)  \citet{Torres2006},
11)  \citet{Kiraga2012},
12) \citet{Garces2011},
13) \citet{Fabricius2002},
14) \citet{Jenkins2009},
15)  \citet{Landolt2009},
16) \citet{vanAltena1995},
17) \citet{Landolt1992}, 
18) \citet{Stauffer1984}\\

\end {table*}
\section{Photometric modulation  analysis}

We search for periodic photometric variability that is compatible with both stellar rotation and long-term magnetic cycles. We compute the power spectrum using a generalised Lomb-Scargle periodogram \citep{Zechmeister2009} and, if there is any significant periodicity, we fit the detected period using a sinusoidal model with the MPFIT routine \citep{Markwardt2009}. Then we repeat the same process in the residuals of the fit. Typically this allowed us to determine  the periodic photometric modulation induced by  the magnetic cycle or the stellar rotation and, in some situations, us to measure both. In those cases in which both quantities have been determined, the final parameters come from a simultaneous fit of both signals. 

The significance of the periodogram peak is evaluated using the \citet{Cumming2004} modification over the \citet{HorneBaliunas1986} formula to obtain the spectral density thresholds for a desired false alarm level. Our false alarm probability then is defined as  $FAP = 1 - [1-P(z > z_{0})]^{M}$, where $P(z > z_{0})$ is the probability of $z,$ the target spectral density,  being greater than $z_{0}$, the measured spectral density, and $M$ is the number of independent frequencies.

As the Nyquist frequency of our data corresponds to a period of two days \citep{Kiraga2007}, we limit our high frequency search to that number, except in a few cases where we have hints that a faster rotation might be happening based on  previous measurements in the literature. On the low frequency side we limit the search to a period of 10000 days, which is almost three times longer than the maximum time span of the observations. The determination of  incomplete cycles obviously involves large associated uncertainties.  

\begin{figure}
\includegraphics[width=9.0cm]{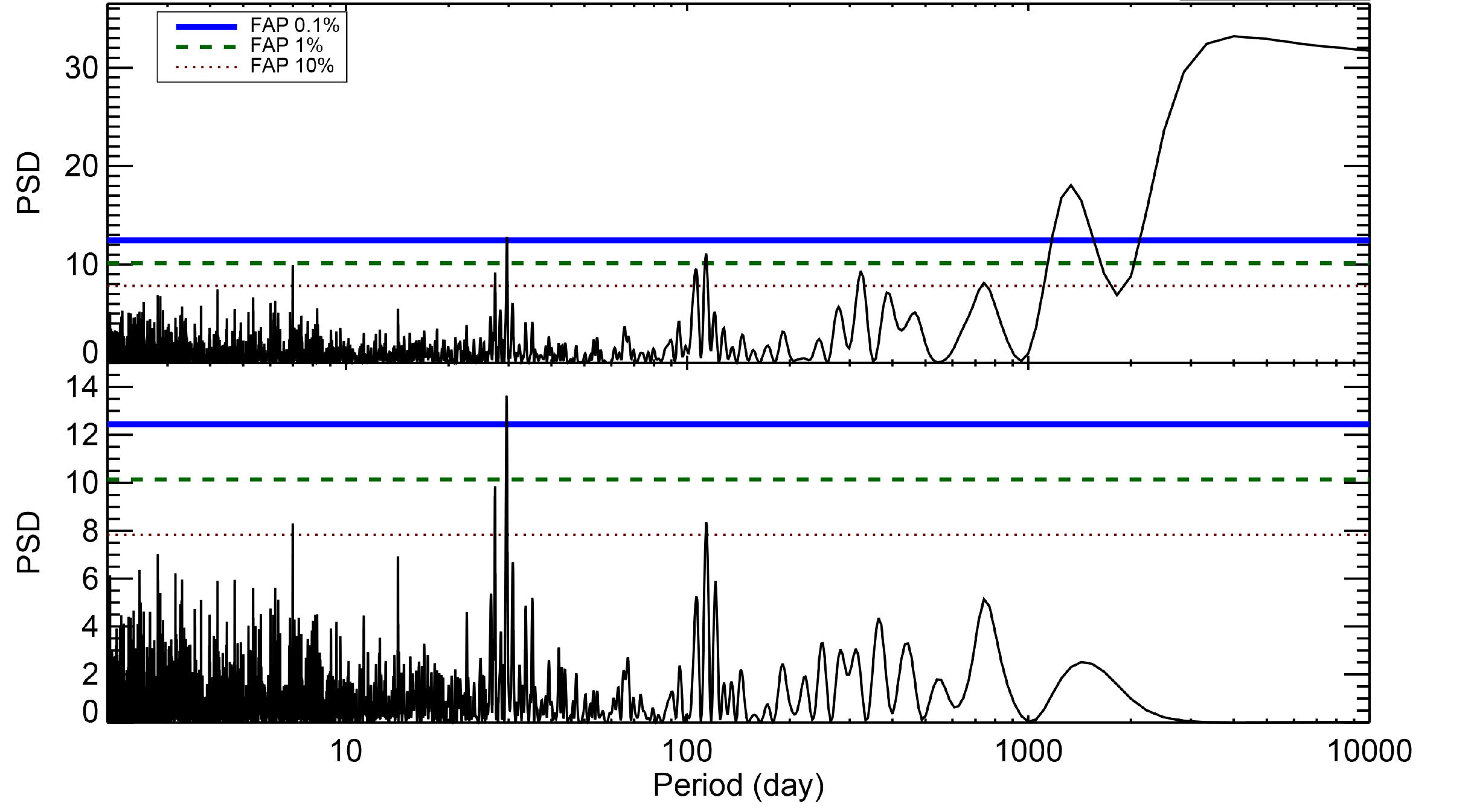}
\caption{Periodograms for the light curve of the star  HD2071. {\it Top panel}:  periodogram of the raw data; {\it bottom panel}:  periodogram of the residuals after fitting the long period signal. }
\label{hd2071_periodograms}
\end{figure}

\begin{figure}
\includegraphics[width=9.0cm]{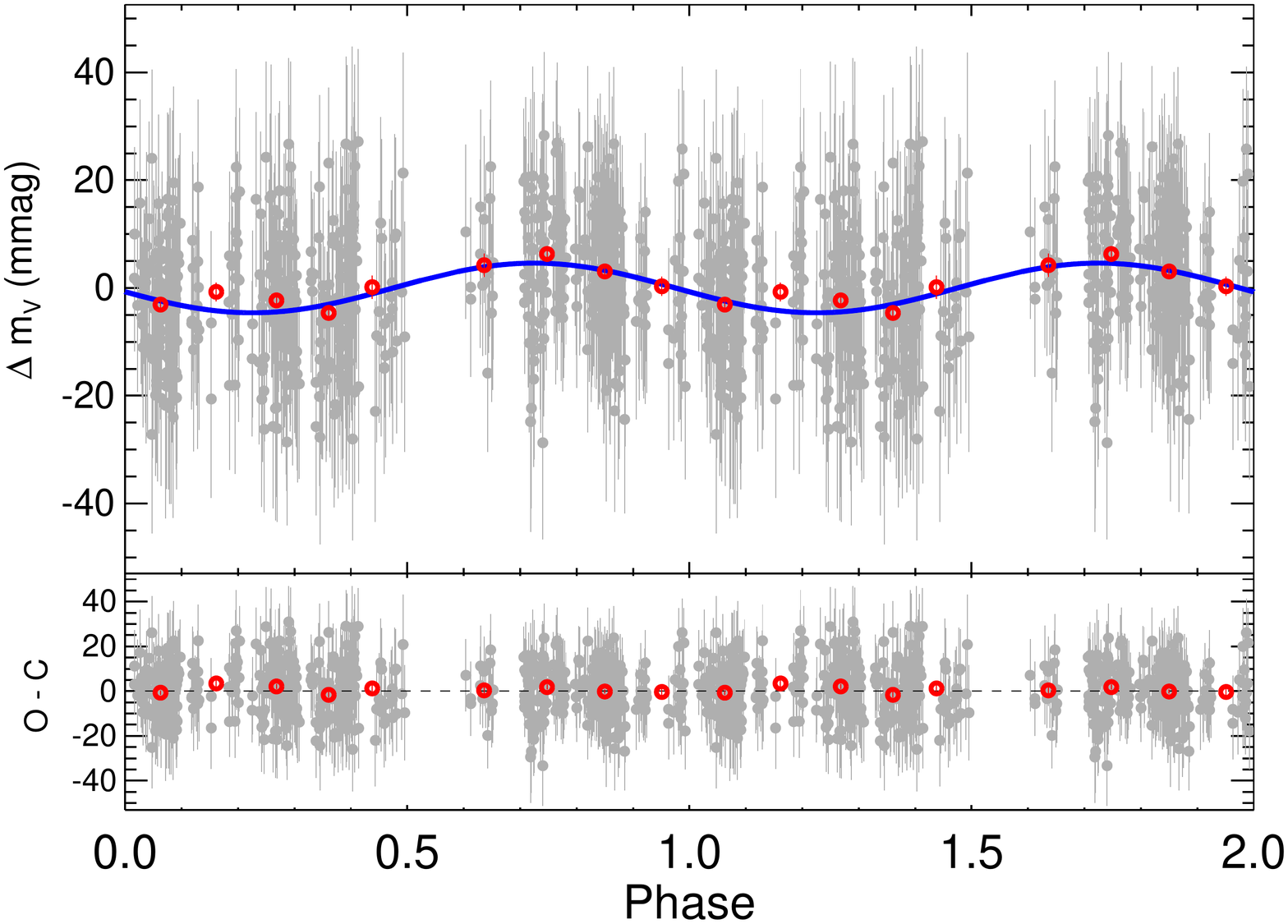}
\caption{Phase fit for the long-term photometric cycle of HD2071. For a period of $11.2 \pm 3.6$ years with a semi-amplitude of $4.9 \pm 1.3\ mmag$. Grey dots show individual measurements, while red dots show the average measurement of every phase bin.  }
\label{hd2071_fits1}
\end{figure}

\begin{figure}
\includegraphics[width=9.0cm]{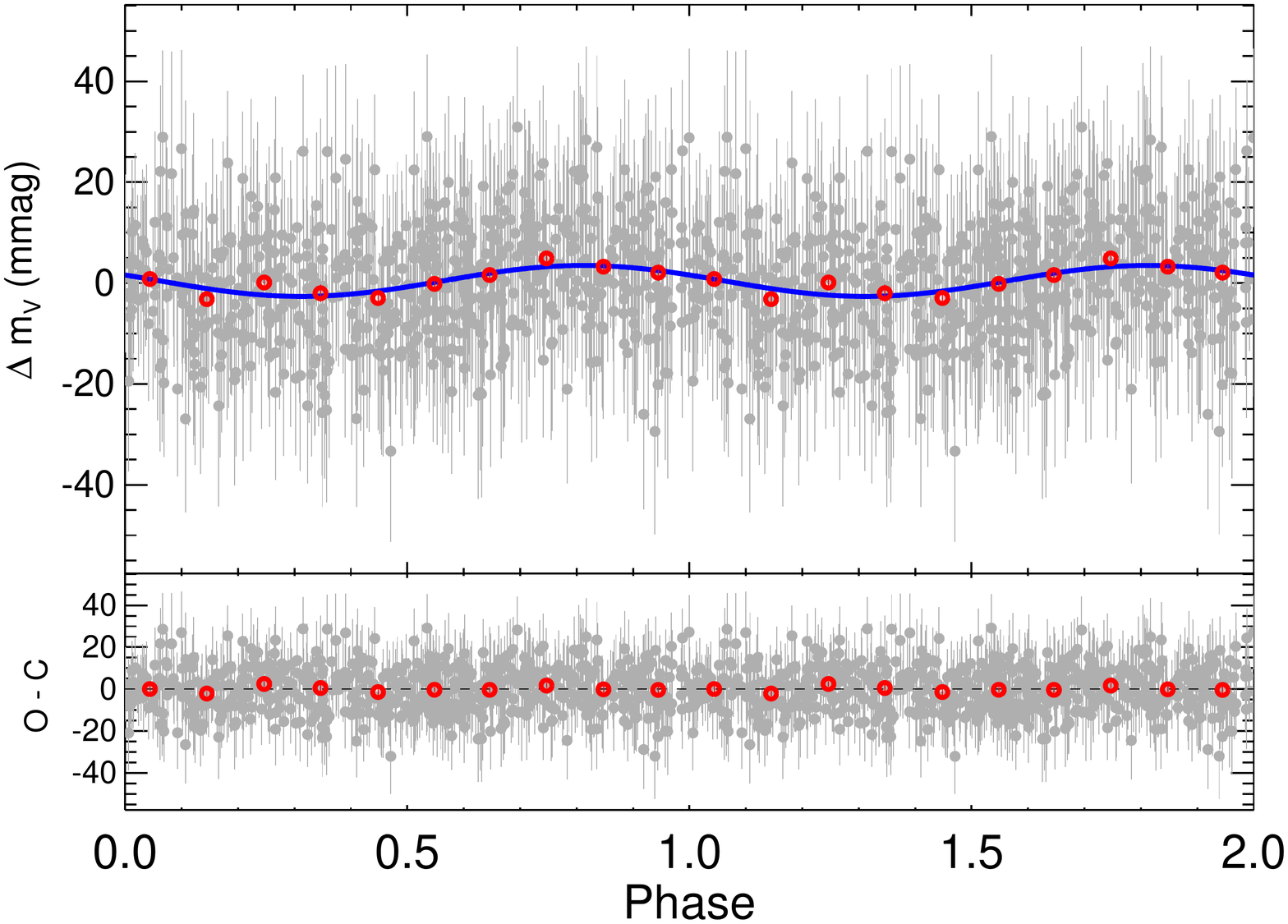}
\caption{Phase fit for the the rotational modulation of HD2071. The measured period is $29.6 \pm 0.1$ days with a semi-amplitude of $3.1 \pm 0.9\ mmag$. Grey dots show individual measurements, and red dots show the average measurement of every phase bin.}
\label{hd2071_fits2}
\end{figure}

In order to illustrate the method,  we describe below the cases  of three stars with  different spectral  types. First, the solar twin HD2071 for which  there are  $641$ photometric  measurements taken over  $8.4$ years. Figure~\ref{hd2071_periodograms} shows the periodograms obtained from the raw data and the residuals after removing the long-term variations. Although there is evidence for a long-term cycle, the data do not cover this cycle completely,  thus  the uncertainty in the  determination of the period of the cycle is large.   Figures~\ref{hd2071_fits1} and ~\ref{hd2071_fits2} show the phase fit for both signals. The best-fitting model includes a long-term cycle of $11.3 \pm 3.6$ years combined with a rotation period of $29.6 \pm 0.1$ days, with semi-amplitudes of $4.9 \pm 1.3\ mmag$ and $3.1 \pm 0.9\ mmag$, respectively. The measured activity cycle of $\sim 11$ years and the rotational period are  similar to the  corresponding cycle and rotation of the Sun.  

\begin{figure}
\includegraphics[width=9.0cm]{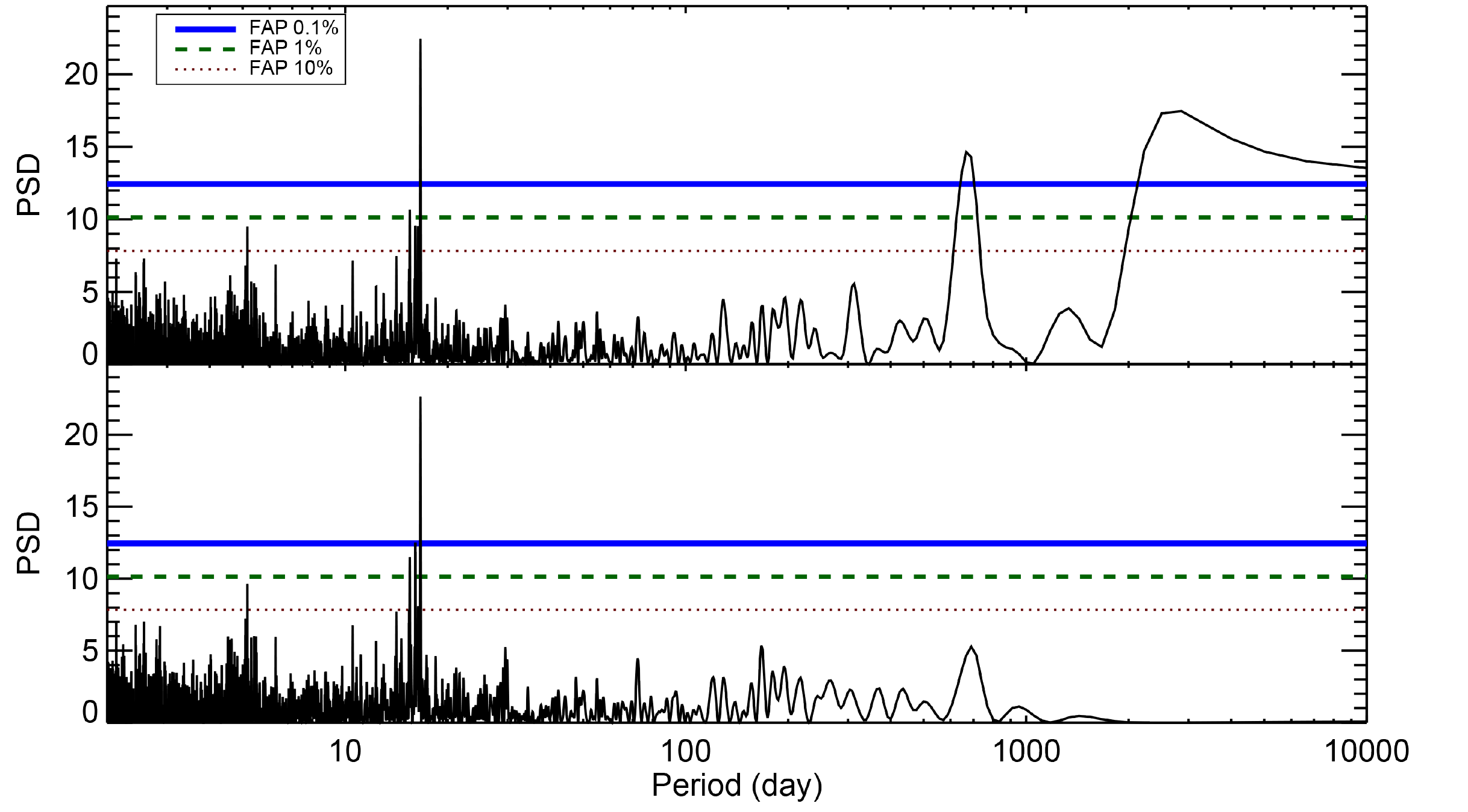}
\caption{Periodograms for the HD224789 light curve. {\it Top panel} shows the periodogram of the raw data; {\it bottom pannel} indicates the periodogram of the residuals after fitting the long period signal. }
\label{hd224789_periodograms}
\end{figure}

\begin{figure}
\includegraphics[width=9.0cm]{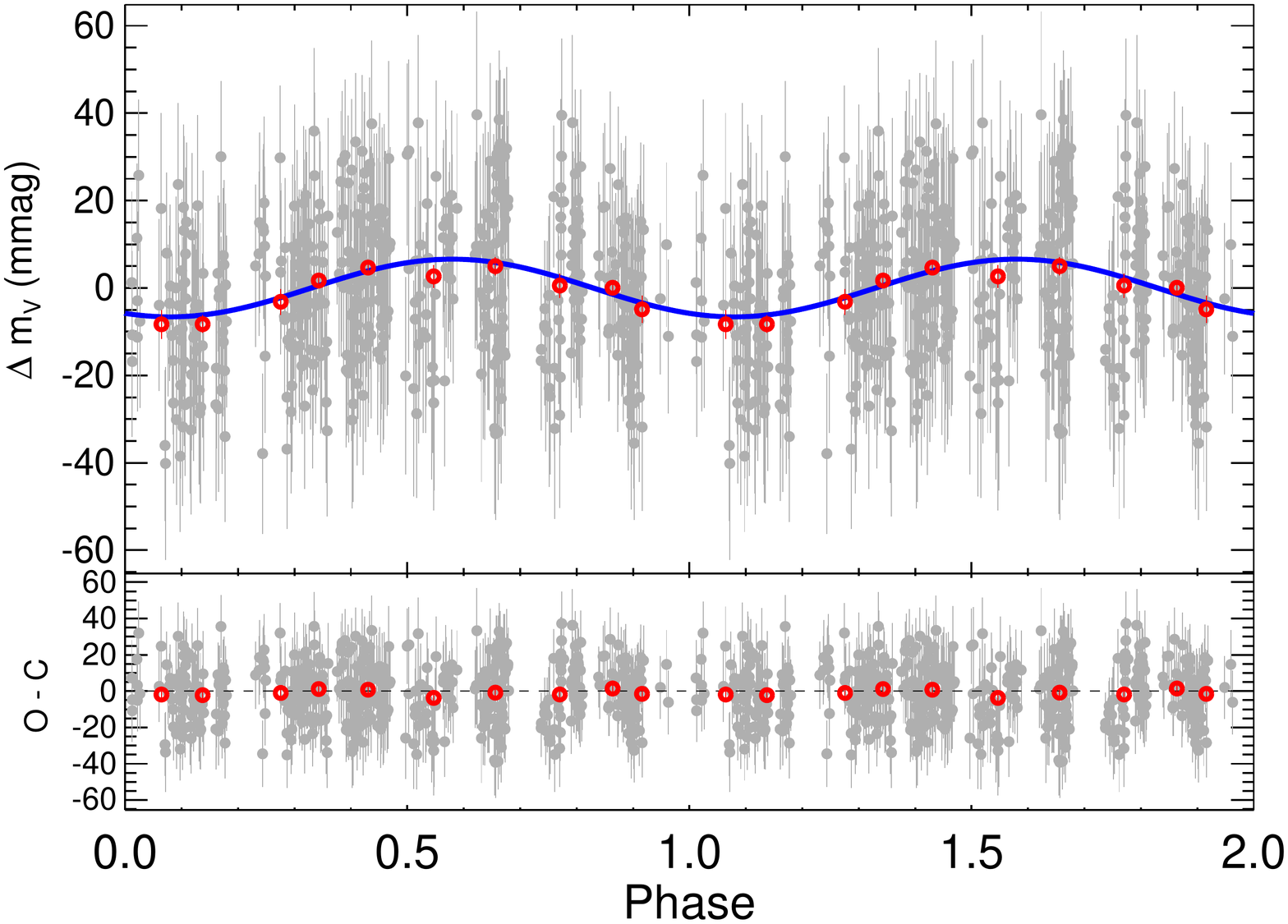}
\caption{Phase fit for the long-term photometric cycle of HD224789. For a period of $7.1 \pm 0.9$ years with a semi-amplitude of $6.7 \pm 1.3\ mmag$. Grey dots show individual measurements, while red ones the average measurement of every phase bin.}
\label{hd224789_fits1}
\end{figure}

\begin{figure}
\includegraphics[width=9.0cm]{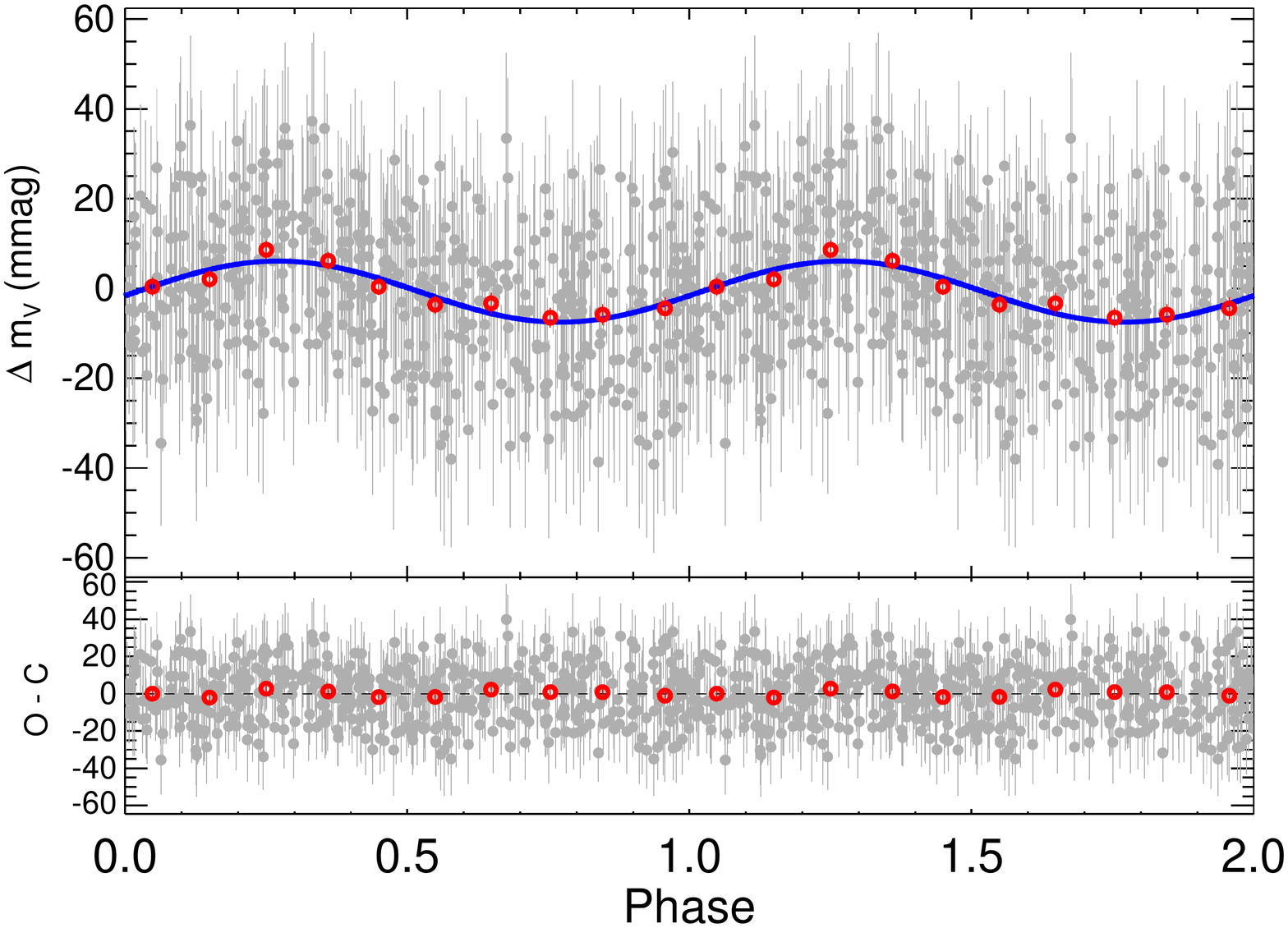}
\caption{Phase fit for the the rotational modulation of HD224789. Measured period is $16.6 \pm 0.1$ days with a semi-amplitude of $6.8\pm 1.2$ mmag. Grey dots show individual measurements, while red ones the average measurement of every phase bin.}
\label{hd224789_fits2}
\end{figure}

For the K-type star HD224789 there are  $489$ measurements obtained along $9.0$ years. In Figure ~\ref{hd224789_periodograms} we plot the corresponding periodograms. We find three significant peaks. The shortest period corresponds to the stellar rotation, the  longest period is likely induced by the  long-term activity  cycle, and the middle period is likely  an alias of the long-term cycle. This  becomes clear  when we fit the longest period signal to the light curve and subtract it;  the intermediate period signal  disappears while the shortest period remains stable. Figures~\ref{hd224789_fits1} and ~\ref{hd224789_fits2} show the phase fit for the two bona fide  signals. The best model includes a long-term cycle of $7.4 \pm 0.8$ years combined with a rotation period of $16.6 \pm 0.1$ days, with photometric  semi-amplitudes of $6.7 \pm 1.3\ mmag$ and $6.8 \pm 1.2\ mmag$, respectively. 

\begin{figure}
\includegraphics[width=9.0cm]{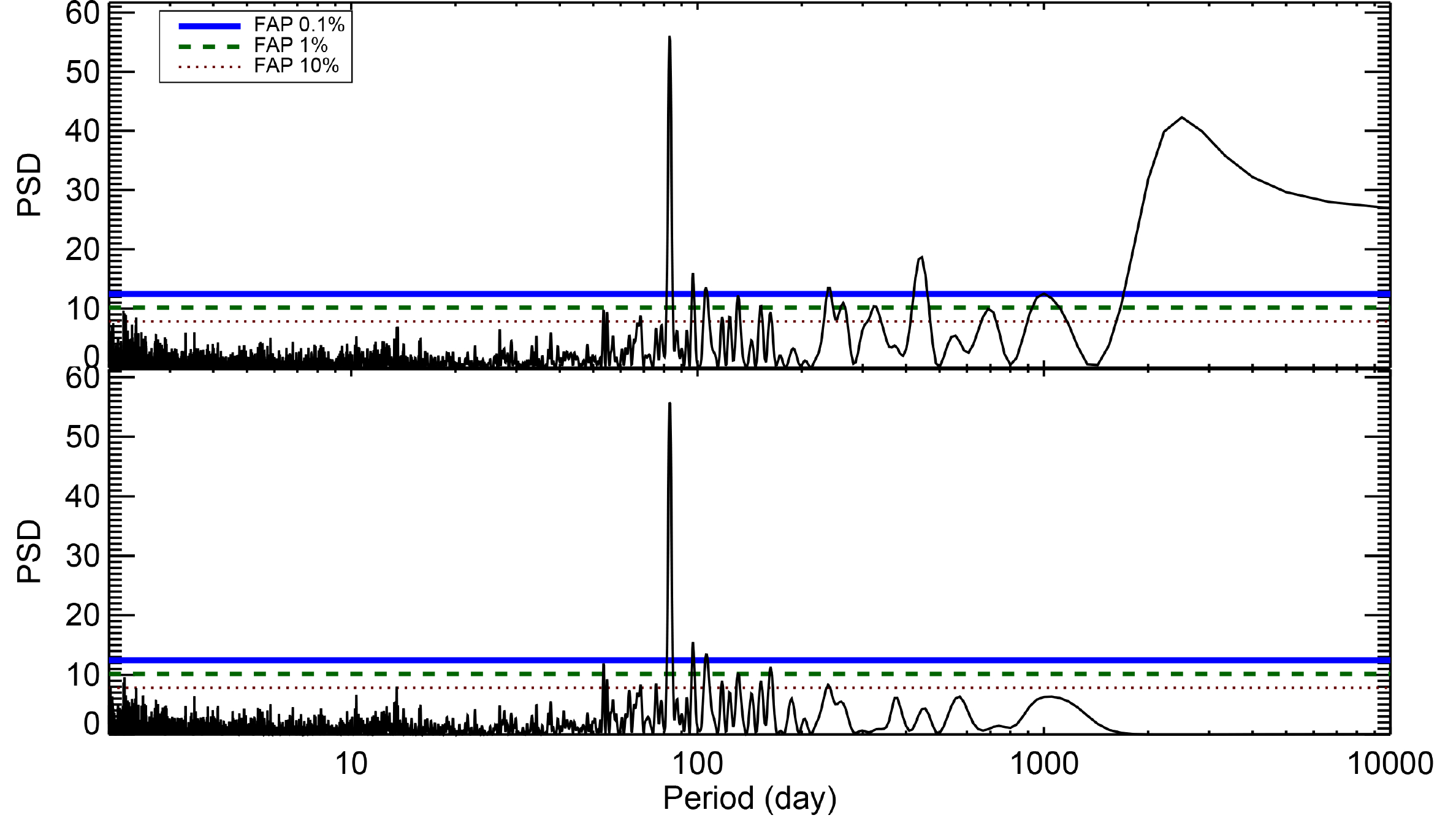}
\caption{Periodograms for the GJ551 light curve. {\it Top panel} shows the periodogram of the raw data; {bottom panel} shows the periodogram of the residuals after fitting the long period signal. }
\label{gj551_periodograms}
\end{figure}

\begin{figure}
\includegraphics[width=9.0cm]{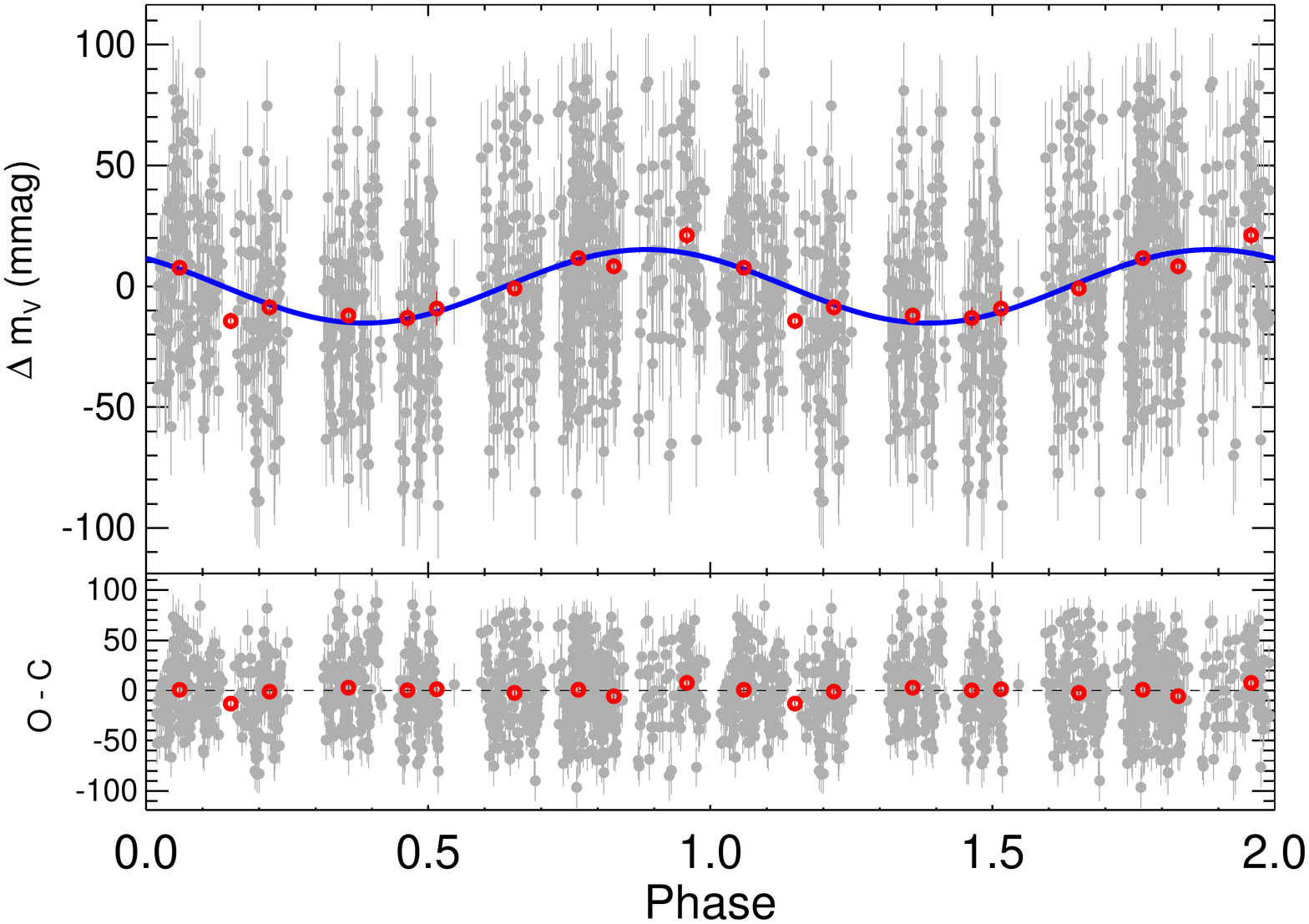}
\caption{Phase fit for the long-term photometric cycle of GJ551. For a period of $6.8 \pm 0.3$ years with a semi-amplitude of $15.5 \pm 0.9$ mmag. Grey dots show individual measurements, while red dots indicate the average measurement of every phase bin.}
\label{gj551_fits1}
\end{figure}

\begin{figure}
\includegraphics[width=9.0cm]{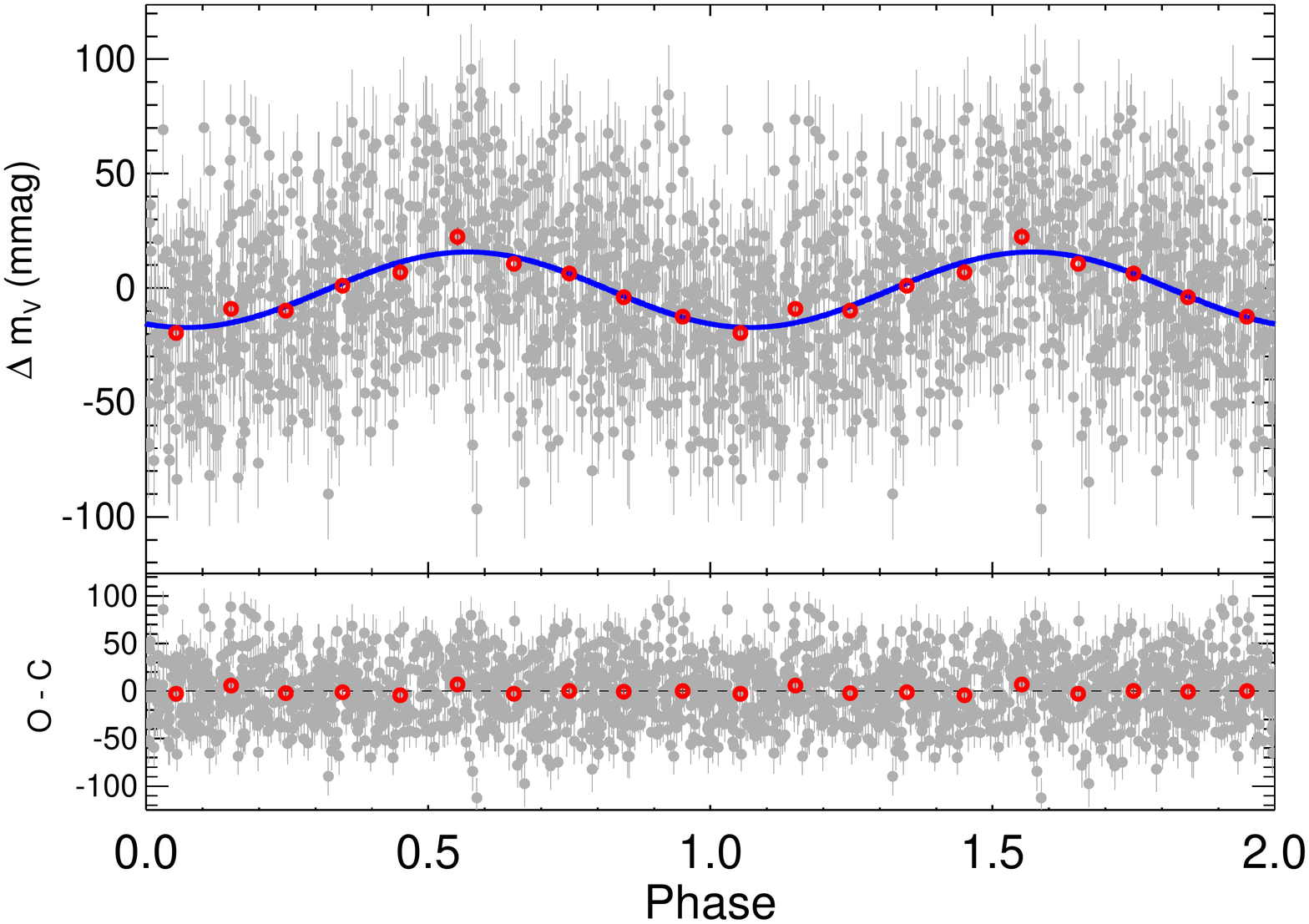}
\caption{Phase fit for  the rotational modulation of GJ551. Measured period is $83.2 \pm 0.1$ days with a semi-amplitude of $16.5 \pm 0.9\ mmag$. Grey dots show individual measurements, while red dots indicate the average measurement of every phase bin.}
\label{gj551_fits2}
\end{figure}

In Fig.~\ref{gj551_periodograms} and Figs.~\ref{gj551_fits1} and ~\ref{gj551_fits2} we show the analysis of the M-type star GJ551 (Proxima Centauri). For this star there are available  $991$ measurements spanning $8.7$ years. We interpret the two obvious   peaks of the periodogram as induced by  the rotation and activity  cycle, respectively. Even if the short period signal is more significant than the long period signal, we opted to fit the long period signal in the first place. The best results comes from a long-term cycle of $6.8 \pm 0.2$ years combined with a rotation period of $83.2 \pm 0.1$ days, with semi-amplitudes of $15.5 \pm 0.9\ mmag$ and $16.5 \pm 0.9\ mmag$, respectively.

Using this procedure we were able to determine  long-term magnetic cycles and/or  average rotation periods for 55 stars in the sample. Table~\ref{Periods} provides the measured cycle and rotation photometric semi-amplitudes and their corresponding periods and false alarm probabilities. The study of variations of the rotational periods across the cycles associated with differential rotation and  the changes in the latitude of spots is out of the scope of the present paper and will be addressed in a future work. The reported rotation periods should be considered  average values over the reported cycles.

\begin{figure}
\includegraphics[width=9.0cm]{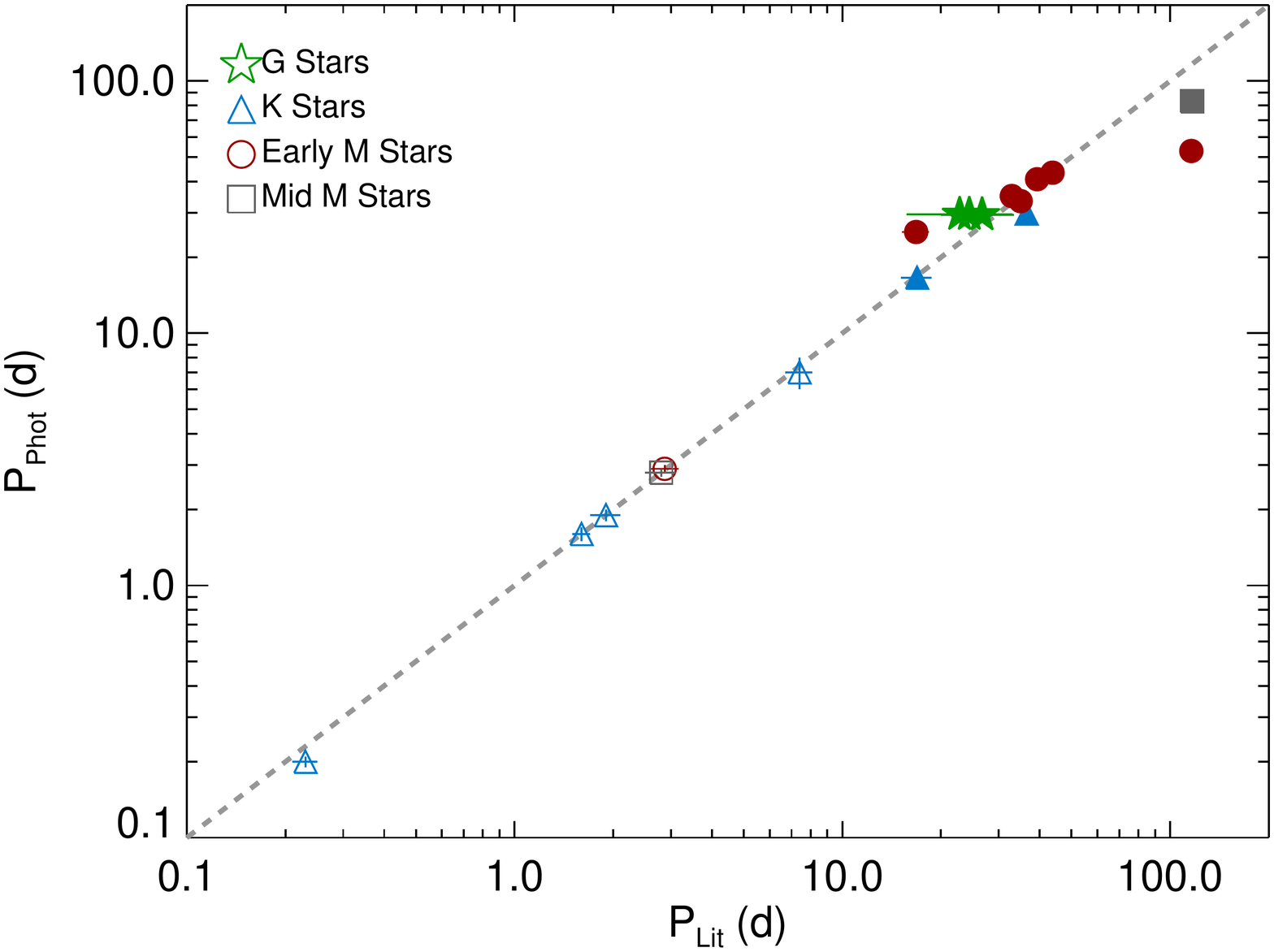}
\caption{Comparison between our measured photometric periods and those reported in the literature from photometric or spectroscopic time series. Filled dots indicate  measurements from  \citet{Masca2015}, while red empty dots are data taken  from  \citet{Chugainov1974}, \citet{Strassmeier1993}, \citet{Kiraga2007}, \citet{Strassmeier2009}, and \citet{Wright2011}. The grey dashed line shows the 1:1 relation.}
\label{rot_rot}
\end{figure}

Figure~\ref{rot_rot} shows the comparison between our measured rotation periods with those previously reported  in the literature. We opt to include only those directly measured using periodic photometric or spectroscopic variations and do not consider  those estimated from activity-rotation relationships. The plot shows good agreement between the different measurements except in the case of one star, GJ 877, where \citet{Masca2015} found roughly twice the rotation period using the variability of the Ca II index. Further study on this star is needed to clarify the correct measurement of the rotation period. It is important to note that the longer a rotation period the harder it is to determine  and  the larger the associated  uncertainties.

\begin{table*}
\begin{center}
\caption {Long-term cycles and rotation periods for the stars in our sample\label{tab:Periods}}
        \begin{tabular}{  l  l  l l l l l l l l l l} \hline 

Star   & $P_{cycle}$ & $A_{cycle}$  & $FAP $& $P_{rot}$ & $A_{rot}$ &$FAP $ & $P_{rot}\ Lit$ & Ref.\\ 
         & $(years)$    &       $(mmag)$        & $(\%)$        &$(d)$  &       $(mmag)$ &$(\%)$  & $(d)$\\\hline
HD1388 & 13.4  $\pm$  6.0 &   8.1  $\pm$  2.2  & $\textless$ 0.1 &  &    & & 19.9 $\pm$ 1.4 & 1\\  
& &     & & &    & & 19.7 $\pm$ 2.8  & 4\\ 
HD10180  & 4.1  $\pm$  0.4 &   3.3  $\pm$  0.9 & 1.7 &  29.5 $\pm$   0.1 &   3.3  $\pm$  0.9 & 3.7 & 24.1 $\pm$ 3.0 & 4\\  
HD21019 &  & &  & 29.5 $\pm$   0.1 &   5.9  $\pm$  1.3 & 0.7 & 38.4 $\pm$ 3.5 & 4\\  

HD2071   & 11.2  $\pm$  3.6 &   4.9  $\pm$  1.3  & $\textless$ 0.1&  29.6 $\pm$   0.1 &   3.1  $\pm$  0.9 & $\textless$ 0.1 & 22.8 $\pm$ 0.7 & 1\\  
HD1320 &  3.3 $\pm$ 0.2 &   2.6  $\pm$  0.8  & 2.1 &  29.6 $\pm$   0.1 &   2.5  $\pm$  0.8& 2.7 & 28.4 $\pm$ 8.7 & 1\\  
HD63765 &  12.9  $\pm$  4.9 &   3.9  $\pm$  0.9   & $\textless$ 0.1 &  29.5 $\pm$   0.1 &   2.7  $\pm$  1.0 & 0.2 & 26.7 $\pm$ 6.7 & 1\\  
HD82558* &  2.5  $\pm$  0.1 &   13.6  $\pm$  1.1  & 0.3&   1.6 $\pm$   0.1 &   28.7  $\pm$  0.9  & $\textless$ 0.1 & 1.60 $\pm$ 0.0 & 6\\  
HD155885   & 6.2 $\pm$  0.1 &   36.9  $\pm$  1.0  &    $\textless$ 0.1 & &    & & 21.1$\pm$ 2.1$^{a}$ & 7\\ 
HD224789 &  7.1  $\pm$  0.9 &   6.7  $\pm$  1.3   & $\textless$ 0.1 &  16.6 $\pm$   0.1 &   6.8  $\pm$  1.2  & $\textless$ 0.1 & 16.9 $\pm$ 1.8 & 1\\  
&   &   & & &   & &   11.8 $\pm$ 4.7  & 4\\  
V660 Tau   & 7.1  $\pm$  0.2 &   31.7  $\pm$  1.5   & $\textless$ 0.1 &  0.2 $\pm$   0.0 &   35.9  $\pm$  1.5  & $\textless$ 0.1 & 0.23 $\pm$ 0.02$^{a}$ & 8\\  
HD176986    & 7.3  $\pm$  0.7 &   4.8  $\pm$  1.2   & $\textless$ 0.1 &   &    & & 33.4 $\pm$ 0.2 & 1 \\  
& &     & & &  & &   39.0 $\pm$ 5.3  & 4\\  
HD32147    & 5.8  $\pm$  0.5 &   7.5  $\pm$  1.3   & $\textless$ 0.1 &   &      & & 47.4 $\pm$ 4.7$^{a}$ & 10\\ 
HD45088*   &  &      & &  7.0 $\pm$   0.1 &   14.2  $\pm$  1.2  & $\textless$ 0.1 & 7.4 $\pm$ 0.7$^{a}$ & 9\\  
HD104067    & 7.7  $\pm$  1.3 &   4.1  $\pm$  0.7  & $\textless$ 0.1 &  27.4 $\pm$   0.1 &   2.8  $\pm$  0.7  & 5.0 & 29.8 $\pm$ 3.1 & 1\\ 

HD215152*    &  & & &  29.8 $\pm$   0.1 &   2.4  $\pm$  1.1 & 3.0 & 36.5 $\pm$ 1.6 & 1 \\ 
& &  &    & & & &    41.8 $\pm$ 5.6 & 4\\  
V410 Tau   &  && & 1.9 $\pm$   0.1 &   253.2  $\pm$  2.7  & $\textless$ 0.1 & 1.9  $\pm$ 0.2$^{a}$ & 12\\ 
HD131977    & 5.1  $\pm$  0.2 &   34.5  $\pm$  1.8   & $\textless$ 0.1 &   &      \\   
 
HD125595    & 5.9  $\pm$  0.6  &   4.1  $\pm$  0.9   & $\textless$ 0.1 &  29.4 $\pm$   0.1 &   2.9  $\pm$  0.9 & 10.3& 37.2 $\pm$ 2.0 & 1\\ 
HD118100     & 9.3 $\pm$ 0.4 & 17.3 $\pm$ 1.1 & $\textless$ 0.1 & 3.9 $\pm$ 0.1 & 30.4 $\pm$ 0.9 & $\textless$ 0.1 \\
GJ9482   & 10.1  $\pm$  1.4 &   6.9  $\pm$  0.9  & $\textless$ 0.1 &   &    \\  
HD113538   & 6.6  $\pm$  0.6 &   4.9  $\pm$  0.9   & $\textless$ 0.1 &  43.9 $\pm$   0.2 &   4.6  $\pm$  0.8 & 2.5 \\  

GJ846   & 8.5  $\pm$  1.1 &   7.3  $\pm$  1.2   & $\textless$ 0.1 &   &     & &31.0 $\pm$ 0.1 & 1\\  
GJ676A    & 7.5  $\pm$  0.6 &   8.2  $\pm$  1.0   & $\textless$ 0.1 &   &     & & 41.2 $\pm$ 3.8 & 1 \\  
GJ229   & 8.4  $\pm$  0.3 &   12.8  $\pm$  0.9  & $\textless$ 0.1  &  27.3 $\pm$   0.1 &   2.9  $\pm$  0.9  & 11.4 \\ 
HD197481   & 7.6  $\pm$  0.4 &   14.4  $\pm$  1.7   & $\textless$ 0.1 &  4.9 $\pm$   0.1 &   18.6  $\pm$  1.1  & $\textless$ 0.1 \\  
GJ514   & 9.9  $\pm$  3.8 &   5.8  $\pm$  1.4   & $\textless$ 0.1 &   &    & & 28.0 $\pm$ 2.9 & 1  \\  
GJ239   & 4.9   $\pm$ 0.4 & 5.2 $\pm$ 1.0 & $\textless$ 0.1\\
GJ205    & 10.8  $\pm$  1.1 &   9.3  $\pm$  0.8   & $\textless$ 0.1 &  33.4 $\pm$   0.1 &   4.8  $\pm$  0.8  & $\textless$ 0.1 & 35.0 $\pm$ 0.1 & 1\\  
                & 3.9   $\pm$ 0.3 &  6.9 $\pm$ 1.4 & $\textless$ 0.1\\
GJ832 &   13.2  $\pm$  4.2 &   6.8  $\pm$  1.5   & $\textless$ 0.1 &   &     & & 45.7 $\pm$ 9.3 & 1 \\  
GJ536    &  & &  & 43.3 $\pm$   0.1 &   4.8  $\pm$  1.1  & $\textless$ 0.1 & 43.8 $\pm$ 0.1 & 1\\  
GJ382   & 13.6 $\pm$ 1.7 & 15.1 $\pm$ 1.0   & $\textless$ 0.1  &  21.2 $\pm$   0.1 &   5.7  $\pm$  0.9  & $\textless$ 0.1 & 21.7 $\pm$ 0.1 & 1\\  
& &     & & &     & & 21.6 $\pm$ 2.2$^{a}$  & 2\\ 
GJ588    & 5.2  $\pm$  0.7 &   3.1  $\pm$  1.2 & 0.5 &  43.1 $\pm$   0.2 &   2.7  $\pm$  1.1 & 2.3 & 61.3 $\pm$ 6.5 & 1\\  
GJ176    & 5.9  $\pm$  0.7 &   6.5  $\pm$  1.6  & 1.5 &  40.8 $\pm$   0.1 &   7.0  $\pm$  1.3 & $\textless$ 0.1 & 39.3 $\pm$ 0.1 & 1 \\  
& &     & & &  & &   38.9 $\pm$ 3.9$^{a}$  & 2\\ 
GJ752A    & 9.3  $\pm$  1.9 &   7.7  $\pm$  3.2   & $\textless$ 0.1 &  46.0 $\pm$   0.2 &   4.5  $\pm$  1.1  & $\textless$ 0.1 & 46.5 $\pm$ 0.3 & 1 \\  
GJ674  & 10.1  $\pm$  0.8 &   17.7  $\pm$  1.2   & $\textless$ 0.1 &  35.0 $\pm$   0.1 &   5.5  $\pm$  1.1  & $\textless$ 0.1 & 32.9 $\pm$ 0.1 & 1\\  
& &     & & &  &&   33.3 $\pm$ 3.3$^{a}$ & 2\\  
GJ680   & 5.0  $\pm$  0.2 &   8.4  $\pm$  0.9   & $\textless$ 0.1 &   &      \\  
GJ408    & 5.3  $\pm$  0.6 &   6.4  $\pm$  1.5 & 2\\
GJ877    &  & & &  52.8 $\pm$   0.1 &   7.8  $\pm$  0.7  & $\textless$ 0.1 & 116.1 $\pm$ 0.7 & 1\\ 
GJ479   & 2.0  $\pm$ 0.1 & 5.7 $\pm$ 1.1 &  $\textless$ 0.1 & 22.5 $\pm$ 0.1 & 4.9 $\pm$ 1.1 & 1.2\\
GJ358*   & 4.9  $\pm$  0.1 &   20.5  $\pm$  0.9   & $\textless$ 0.1 &  25.2 $\pm$   0.1 &   12.1  $\pm$  0.7  & $\textless$ 0.1 & 16.8 $\pm$ 1.6 & 1\\  
& &     & & &    & & 25.0 $\pm$ 2.5$^{a}$  & 2\\ 
GJ273    & 6.6  $\pm$  1.3 &   7.7  $\pm$  1.5   & $\textless$ 0.1&   &    && 115.9 $\pm$ 19.4 & 1  \\  
GJ849   & 10.2  $\pm$  0.9 &   12.4  $\pm$  1.0   & $\textless$ 0.1 &   &    & &  39.2 $\pm$ 6.3 & 1 \\  
GJ729   & 7.1 $\pm$ 0.1 & 5.3 $\pm$ 0.7 &  $\textless$ 0.1 & 2.9 $\pm$ 0.1 & 8.8 $\pm$ 0.9 & $\textless$ 0.1 & 2.87 & 2\\
& 2.1 $\pm$ 0.1 & 7.9 $\pm$ 0.5 &  $\textless$ 0.1 
\\
GJ896A    &  & & &  15.3 $\pm$   0.1 &   25.2  $\pm$  1.3 & $\textless$ 0.1 \\  
GJ317    & 5.2  $\pm$  0.3 &   12.4  $\pm$  1.1 & $\textless$ 0.1 &     \\  
GJ526   & 9.9  $\pm$  2.8 &   6.9  $\pm$  1.0   & $\textless$ 0.1 &   &     & & 52.3 $\pm$ 1.7 & 1 \\  
GJ234   & 5.9  $\pm$  0.5 &   10.1  $\pm$  1.4  & $\textless$ 0.1 &  8.1 $\pm$   0.1 &   7.7  $\pm$  1.3 & 0.3\\  

GJ54.1  &                                 &                                       &              &  69.2  $\pm$  0.1  & 15.6  $\pm$ 1.1 &  $\textless$  0.1\\

\hline
\end{tabular} 
 \end{center}

\end    {table*}

\begin{table*}
\ContinuedFloat
\begin{center}
\caption {Long-term cycles and rotation periods for the analysed stars\label{tab:Periods}}
        \begin{tabular}{ l  l  l l l l l l l l l l} \hline \\

Star &  $P_{cycle}$ & $A_{cycle}$  & $FAP $& $P_{rot}$ & $A_{rot}$ &$FAP $ & $P_{rot}\ Lit$ & Ref.\\ 
        &        $(years)$      &       $(mmag)$        & $(\%)$        &$(d)$  &       $(mmag)$ &$(\%)$  & $(d)$\\\hline
        GJ628    & 4.4  $\pm$  0.2 &   8.3  $\pm$  1.1  & $\textless$ 0.1 &  119.3 $\pm$   0.5 &   5.9  $\pm$  1.0  & $\textless$ 0.1\\ 
LP 816-60    & 10.6  $\pm$  1.7 &   8.2  $\pm$  1.2  & 3.0 &  67.6 $\pm$   0.1 &   10.7  $\pm$  1.1 & $\textless$ 0.1\\ 
GJ285   & 10.6  $\pm$  0.4 &   37.2  $\pm$  1.1 & $\textless$ 0.1  &  2.8 $\pm$   0.1 &   23.2  $\pm$  1.0 & $\textless$ 0.1 & 2.8 $\pm$ 0.3$^{0.3}$ & 11\\  
GJ447   & 4.1  $\pm$  0.3 &   7.1  $\pm$  1.1  & 1.7 &  165.1 $\pm$   0.8 &   7.8  $\pm$  1.1 & 0.2 \\  
GJ581    & 6.2 $\pm$  0.9 &   3.8  $\pm$  0.9   & $\textless$ 0.1 &   &    &  & 132.5 $\pm$ 6.3 & 1 \\  
& &     & & &  &&   130.0 $\pm$ 2.0 & 3\\  
GJ551   & 6.8  $\pm$  0.3 &   15.5  $\pm$  0.9  & $\textless$ 0.1 &  83.2 $\pm$   0.1 &   16.5  $\pm$  0.9 & $\textless$ 0.1 & 116.6 $\pm$ 0.7 & 1\\
& &     & & &  &&   82.5 $\pm$ 8.3$^{a}$  & 2\\  
GJ406   & 8.9 $\pm$ 0.2 & 64.8 $\pm$ 1.2 & $\textless$ 0.1 \\
 \hline
\label{Periods}

\end{tabular} 
 \end{center}
 
* Long-term linear trend. \\ 
$^{a}$ No uncertainty provided. 10 per cent of the measured period adopted as standard uncertainty.\\
References for rotation periods in the literature: 
1) \citet{Masca2015} using periodic variations of $Ca_{II}\ H \& K$ and $H_{\alpha}$ emission lines.
2) \citet{Kiraga2007} using periodic variations in the photometry of the ASAS survey.
3) \citet{Robertson2014} using periodic variations of $H_{\alpha}$ emission line. 
4)  \citet{Lovis2011} using \citet{Mamajek2008} age-activity-rotation relationship.
5) \citet{OlmedoChavez2013} using their own activity-rotation relationship based on the $Mg_{II}\ H \& K$ emission lines.
6) \citet{Jetsu1993} using periodic variations in photometry.
7) \citet{Donahue1996} using periodic variations in photometry.
8) \citet{Wright2011} using periodic variations in photometry.
9) \citet{Strassmeier1993} using periodic variations in photometry.
10)  \citet{Saar1997B} estimated from $Ca_{II}\ H \& K$ measurements.
11) \citet{Chugainov1974} using periodic variations in brightness.
12) \citet{Strassmeier2009} using periodic variations in photometry.\\
\end    {table*}

For the remaining 70 stars  it was not possible to achieve  a determination of rotation periods or activity cycles.   In spite of  the quality control on the data, some of these stars were so bright that their  measurements where frequently  in the non-linear regime and thus  provided unreliable light curves.  In some other stars, the amplitude of the induced photometric variability is below our detection limit or is non-periodic. 

\subsection{Long-term linear trends}

Four of our stars, i.e. HD82558, HD45088, HD215152 and GJ358, showed clear long-term linear trends over the time span of our observations. In the case of HD45088 and HD215152, these trends implied the presence of cycles much longer than the time span of the observations.  For HD82558 and GJ358, we also identify long-term trends on top of the  $\sim$ 2.5 yr and $\sim$ 5 yr cycles  that we found for each of these stars. Rotation periods are reported for the four stars. 

Some of the measured cycles are longer than the time span of the observations (HD1388, GJ9482, GJ205, GJ832, GJ382, GJ674, GJ849, and GJ285). While in those cases we have not measured a full cycle yet, we have enough data to give an estimation of the cycle length instead of treating it as a trend.

\subsection{Amplitude modulation}
Magnetic cycles do not only manifest themselves as changes in the brightness of the stars, but some times in variations of the amplitude of the light curve \citep{BerdyuginaJarvinen2005}. To study this behaviour we regroup the light curve in 30 day bins and use the peak-to-peak magnitude of each bin as a measurement of the amplitude. We then analyse the resulting amplitude curves for all the stars of our sample in the same way as the previous light curves searching for periodicities. 

We find that the stars HD32147 and HD113538 show clear modulations, with false alarm probabilities smaller than 0.1\% and periods of 10.17 $\pm$ 2.28 yr and 8.74 $\pm$ 1.24 yr respectively.  The following stars show less significant modulations: GJ588 shows a periodicity of 1.76 $\pm$ 0.09 yr with a FAP of 2.9\%; GJ273 shows a periodicity of 8.67 $\pm$ 2.41 yr with a FAP of 10.6\%; HD197481 shows a periodicity of 2.77 $\pm$ 0.06 yr with a FAP of 16.5\%, and GJ382 shows a periodicity of 7.77 $\pm$ 1.06 yr with a FAP of 39\%. There is also one short period modulation. For the star GJ9482, we find an amplitude periodicity of 133.55 $\pm$ 0.75 days with a FAP of 2.5\%. The origin of this periodicity is hard to assess. It seems to be too long to be the rotation of the star and too short to be a magnetic cycle.

\subsection{Notes on individual stars}

\textbf{HD10180:} \citet{Lovis2011} provides a cycle measurement of 7.4 $\pm$ 1.2 yr by studying the variations in the Ca II  $~H\&K$ emission cores. Instead of that we find a 4.1  $\pm$  0.4 yr variation. Using the HARPS-S data, we verified that  we detect a  $\sim$  3.2 yr cycle on top of a $\sim$  14 yr variation in the Ca II $~H\&K$ emission time series. We cannot rule out that we detect a harmonic of the real cycle. 

\noindent\textbf{HD155885:} \citet{Hempelmann1995} and \citet{SaarBrandenburg1999} proposed cycles ranging from 5 to 6 yr, with large associated uncertainty. Our measurement of a  6.2 $\pm$ 0.1 yr photometric periodicity supports their previous claims. 

\noindent\textbf{HD63765:} While we retrieve a 12.9 $\pm$ 4.9 yr cycle in the photometric light curve, \citet{Lovis2011} found a periodicity of 6.1 $\pm$ 2.8 yr in the Ca II $~H\&K$ emission time series. We did not find a 6 yr periodicity, but this measure is compatible with the first harmonic of our measurement.

\noindent\textbf{HD32147:} \citet{Baliunas1995} found a 11.1 $\pm$ 0.2 yr cycle studying the Ca II $~H\&K$ emission time series. Our 5.8 $\pm$ 0.5 yr measurement is compatible with the first harmonic of that cycle. 

\noindent\textbf{HD82558:} Previous studies have reported different cycle lengths for this star, ranging from $\sim$ 2.5 up to $\sim$ 12 yr \citep{Olah2009}. In our data we recover the $\sim$ 2.5 yr cycle, which might be a flip-flop cycle, and is also evidence for an additional long-term trend. 

\noindent\textbf{GJ205:} \citet{Savanov2012} finds a 3.7 yr period cycle and some of shorter periods. Our best fit comes from the superposition of two cycles, one of 10.8 $\pm$ 1.1 yr and another one of 3.9 $\pm$ 0.3 yr cycle. The short period cycle might be a flip-flop cycle. 

\noindent\textbf{GJ729:} For this star we find two superimposed cycles. One of 7.1 $\pm$ 0.1 yr and a shorter cycle of 2.1 $\pm$ 0.1 yr. Again the short period cycle might be a flip-flop cycle. 

\noindent\textbf{GJ234:} This is close binary system composed of two low mass stars with a brightness difference of $\Delta m_{K}$ $\sim$ 1.6 and a separation of $\sim$ 1 arcsec \citep{Ward-Duong2015}. Although the photometric light curve is for the unresolved binary, the signal is likely dominated by the primary component given its dominance in the visible range. We ascribe the periodic signals detected in the light curve to the primary. 

\noindent\textbf{GJ581:} \citet{Robertson2013} detects a 4.5 $\pm$ 0.3 years cycle analysing the $H_{\alpha}$ time series whereas we detect a 6.2 $\pm$  0.9 years cycle in the photometry light curve.

\noindent\textbf{GJ551:} \citet{Cincunegui2007} claimed a 1.2 yr cycle for which we find no evidence in the photometric light curve.  Instead we measure a clear 6.8  $\pm$  0.3 yr cycle. \citet{Savanov2012} found a 7.9 yr cycle.

\section{Discussion}

The previous analysis of variability in photometric time series  provided a collection of magnetic cycles, rotation periods, and chromospheric activity level for 55 stars, out of which  34 are low activity  M-type stars, for which only a few tens of magnetic cycles are reported  in the literature. 

We measured the magnetic cycles for 5 G-type stars for which we found a mean cycle length of $9.0$ yr with a dispersion of $4.9$ yr.   We found a mean cycle length of $6.7$ yr with a dispersion of 2 years for 12 K-type stars.  We measured a  mean cycle length of $7.4$ yr with a dispersion of $3.0$ yr for 22 early
M-type stars, and we found  a mean cycle length of $7.56$ yr with a dispersion of $2.6$ yr for 9 mid M-type stars. We note that we might be mixing some flip-flop cycles along with the global cycles, but in most cases we do not find multiple cycles making it difficult to distinguish between the two types. Subsequently, we measured the rotation periods for 9 G-type stars with a mean rotation period of $27.9$ days and a dispersion of $3.9$ days. For 13 K-type stars we measured a typical rotation of $20.3$ with a dispersion of  $16.4$ days. For 20 early M-type stars, we measured an average rotation period of $35.7$ days with a dispersion of $23.2$ days,  and an average rotation period of  $78.1$ days with a dispersion of $54.8$ days for 9 mid M-type stars. 

In order to put these results on a broader context  we include in the discussion and plots  other FGKM stars with  known cycles and rotation periods selected from  \citet{Noyes1984}, \citet{Baliunas1995}, \citet{Lovis2011},  \citet{Robertson2013}, and \citet{Masca2015}. In total we deal with more than 150 stars with similar number of G-, K-, and M-type and far fewer  F-type stars to study the distribution of cycle lengths and rotation periods for the different spectral types and the activity-rotation relationships. 

\subsection{Cycle length distribution}

\begin{figure}
\includegraphics[width=9.0cm]{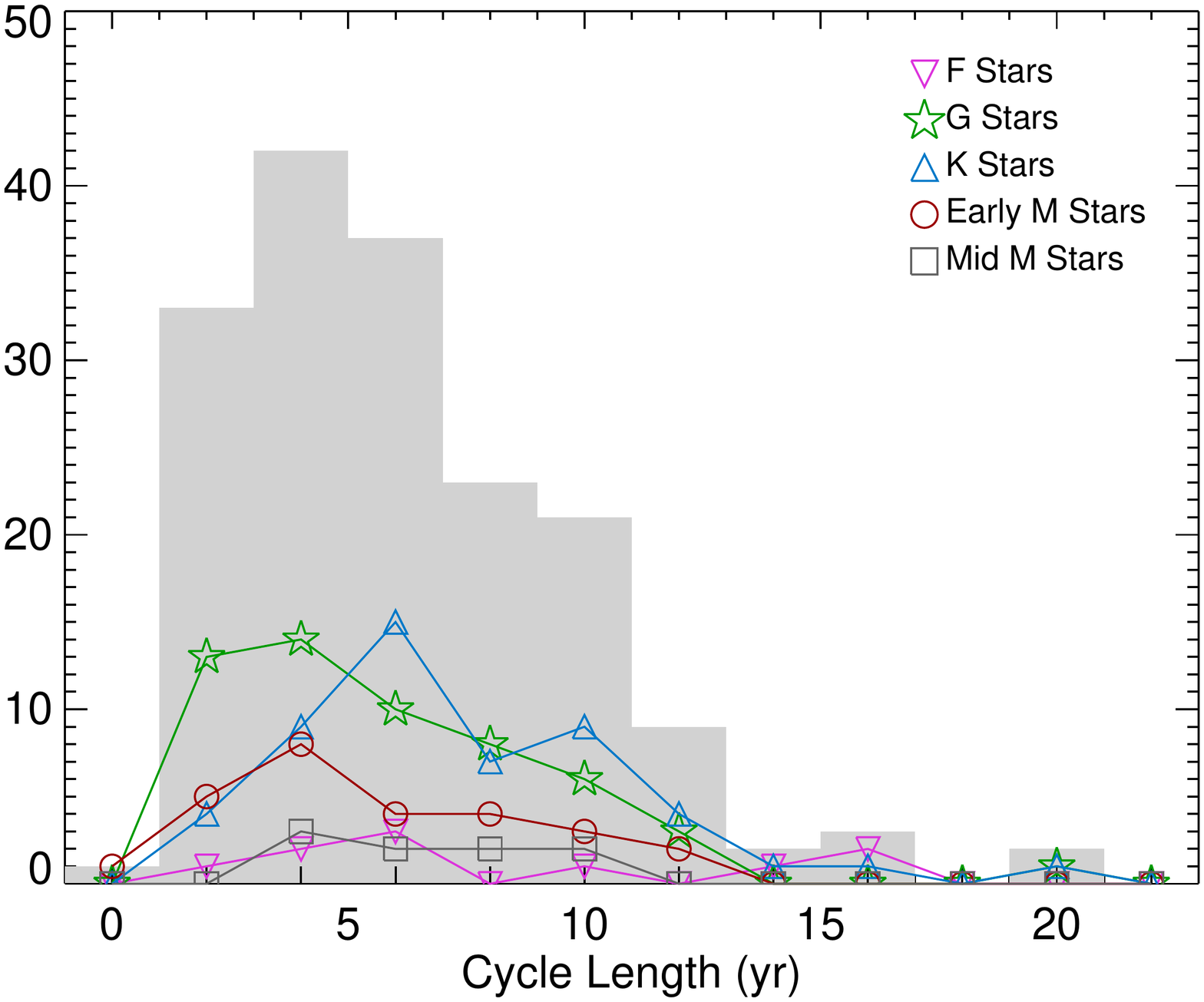}
\caption{Distribution of cycle lengths. Grey filled columns show the global distribution while coloured lines the individual distributions.}
\label{cic_histo}
\end{figure}

Figure~\ref{cic_histo} shows the distribution histogram of cycles by length and spectral type. We find that, like G-type stars, early M-type stars peak at the 2-4 year bins, but K-type stars peak at the 6 year bin. The double peak seen in the distribution might reveal information about peak in global cycles (10 yr) and in flip-flop cycles (6 yr). There are not enough detections  to see any particular behaviour for F-type and mid M-type stars. Table~\ref{cycle_stats} shows the main  statistics of the typical cycle for each spectral type.

\begin {table}
\begin{center}
\caption {Statistics of the length of  known cycles\label{tab:cycle_stats}}
    \begin{tabular}{  l l  l  l l } \hline
Sp. Type  & N &  Mean length & Median length & $\sigma$  \\ 
        &       &  (yr) & (yr) &(yr) \\\hline
F                       & 10 & 9.5  &  7.4 & 5.3 \\
G                       & 55     & 6.7  &  6.0 & 3.6 \\
K                       & 51 & 8.5  &  7.6 & 3.6 \\
Early M                 & 47 & 6.0  &  5.2 & 2.9 \\
Mid M           & 10 & 7.1  &  6.8 & 2.7 \\

 \hline
\label{cycle_stats}
\end{tabular}  
\end{center}

\end {table}

\subsection{Rotation period distribution}

\begin{figure}
\includegraphics[width=9.0cm]{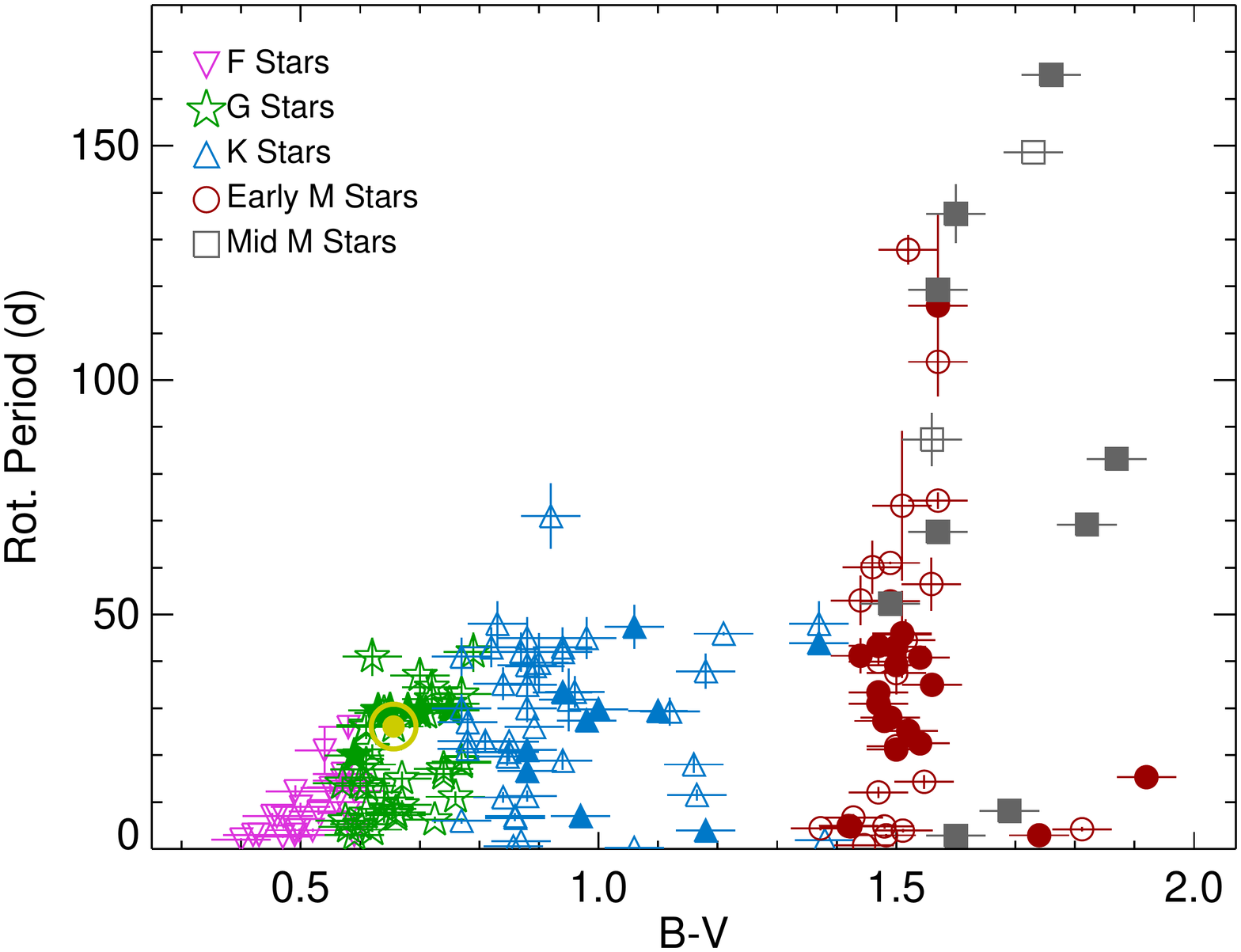}
\caption{Rotation periods vs. B-V colour of stars analysed in this work (filled symbols) and stars from the literature (open symbols). }
\label{rot_histo}
\end{figure}

 Figure~\ref{rot_histo} shows the distribution of rotation periods and Table~\ref{rot_stats} lists the typical periods and measured scatter. When looking at the rotation periods we find an upper limit of the distribution growing steadily towards larger periods for later spectral types, and saturating at $\sim$ 50 days for almost all spectral types (see Fig.~\ref{rot_histo}). M-type stars, especially mid-Ms, show larger scatter going over that saturation limit and reaching periods longer than 150 days.

\begin{figure}
\includegraphics[width=9.0cm]{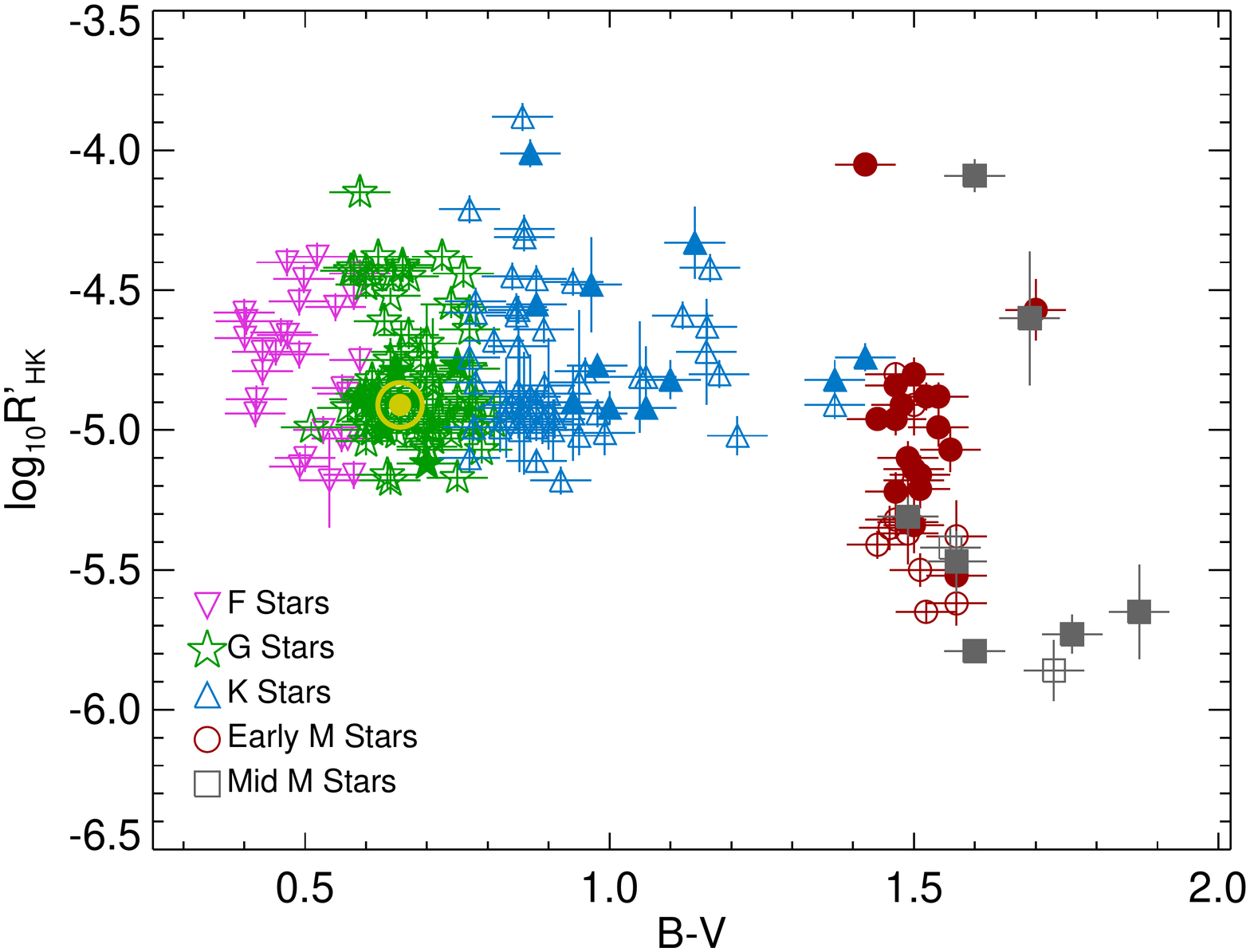}
\caption{Chromospheric activity level $\log_{10}R'_{HK}$ against the colour B-V of the stars. Filled symbols show the stars analysed in this work. }
\label{bv_RHK}
\end{figure}

Many of the M-type stars are extreme slow rotators and that get reflected in their extremely low chromospheric activity levels. Figure~\ref{bv_RHK} shows the distribution of the $\log_{10}R'_{HK}$ against the colour B-V. For solar-type stars the lower envelope of the distribution goes around $\sim -5.2$, but M dwarfs reach level of almost $\sim -6$. 

\begin {table}
\begin{center}
\caption {Statistics of rotation periods\label{tab:rot_stats}}
    \begin{tabular}{  l l  l  l l } \hline
Sp. Type  & N &  Mean Period & Median Period & $\sigma$ Period \\ 
        &       &  (d) & (d) &(d) \\\hline
F                       & 25 & 8.6  &  7.0 & 6.2 \\
G                       & 44     & 19.6  &  18.4 & 11.1 \\
K                       & 53 & 27.4  &  29.3 & 15.7 \\
Early M                 & 43 & 36.2  &  33.4 & 29.9 \\
Mid M           & 11 & 85.4  &  86.2 & 53.4 \\

 \hline
\label{rot_stats}
\end{tabular}  
\end{center}

\end {table}

\subsection{Activity-rotation relation}

\begin{figure*}
\includegraphics[width=20.0cm]{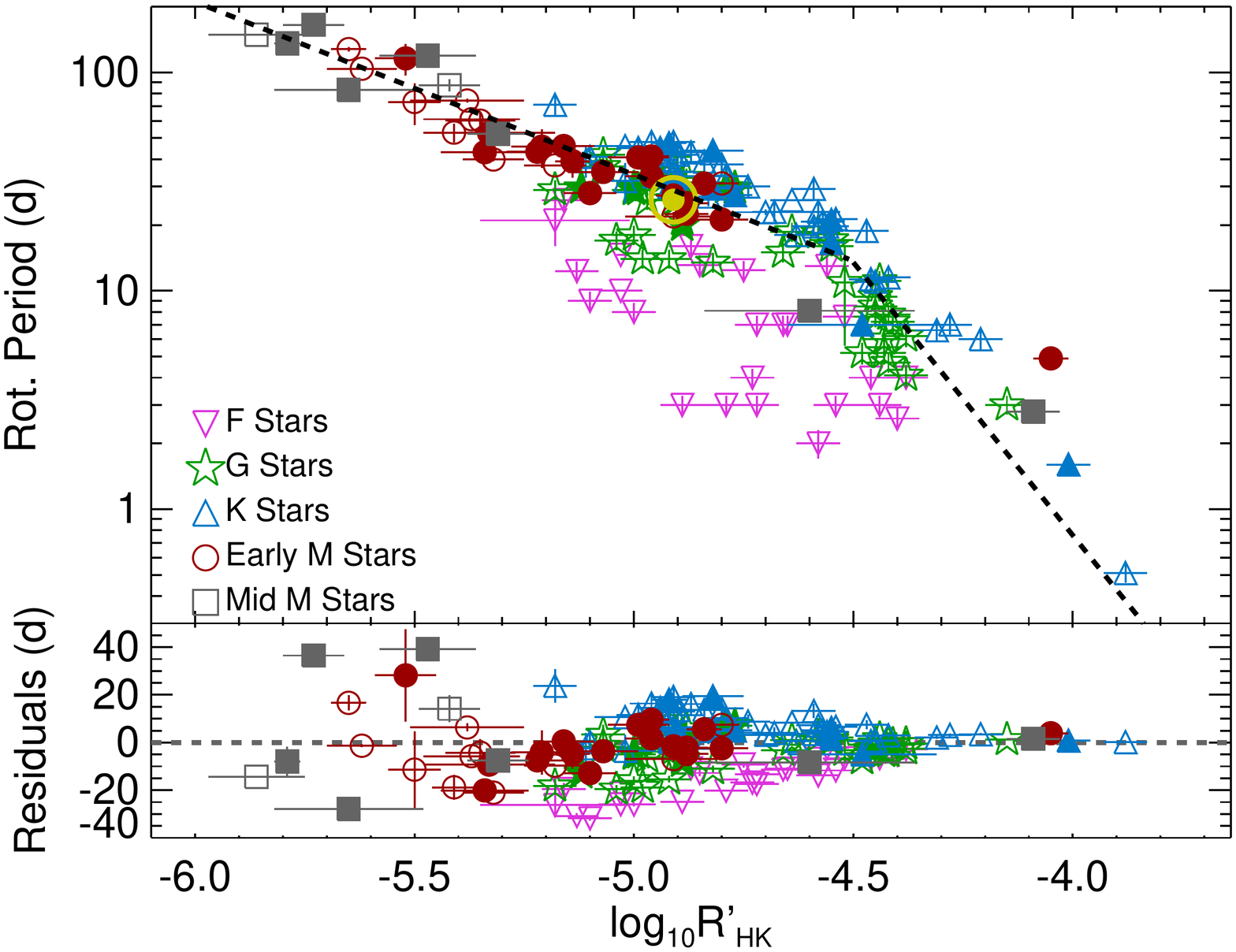}
\caption{Rotation period vs. chromospheric activity level $\log_{10}R'_{HK}$. Filled symbols show the stars analysed in this work. The dashed line shows the best fit to the data, leaving out the F-type stars.}
\label{rhk_period}
\end{figure*}

\begin{figure}
\includegraphics[width=9.0cm]{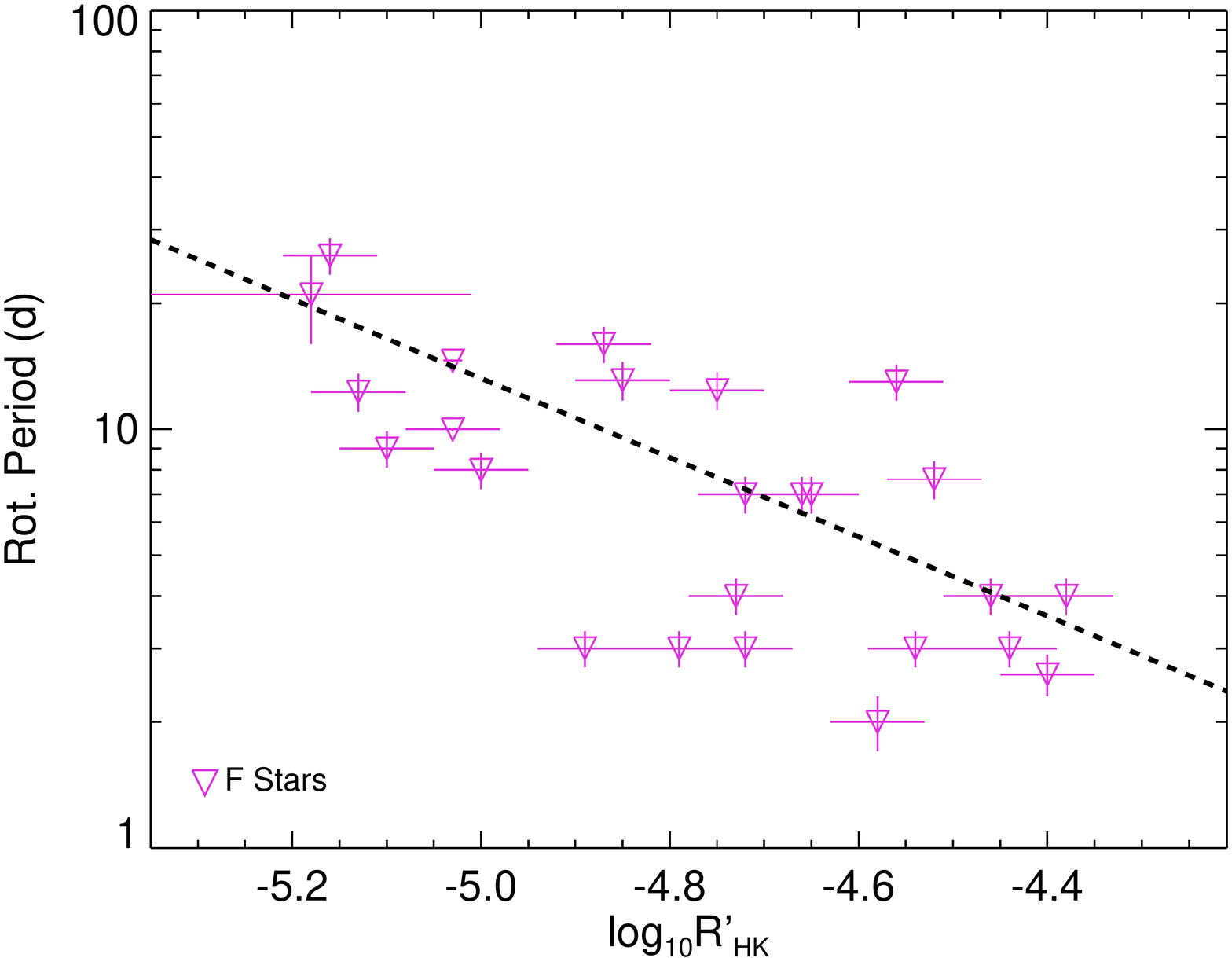}
\includegraphics[width=9.0cm]{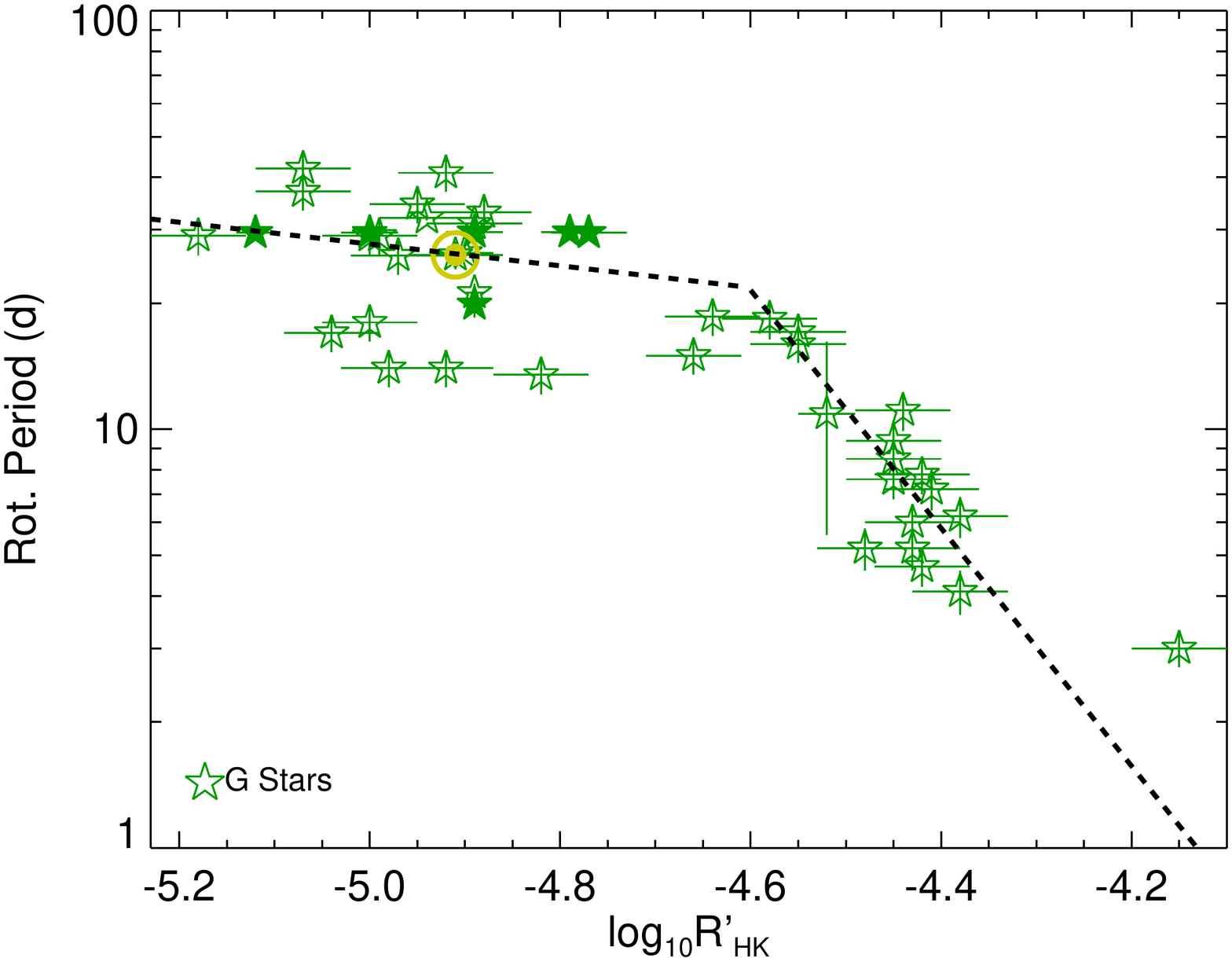}
\includegraphics[width=9.0cm]{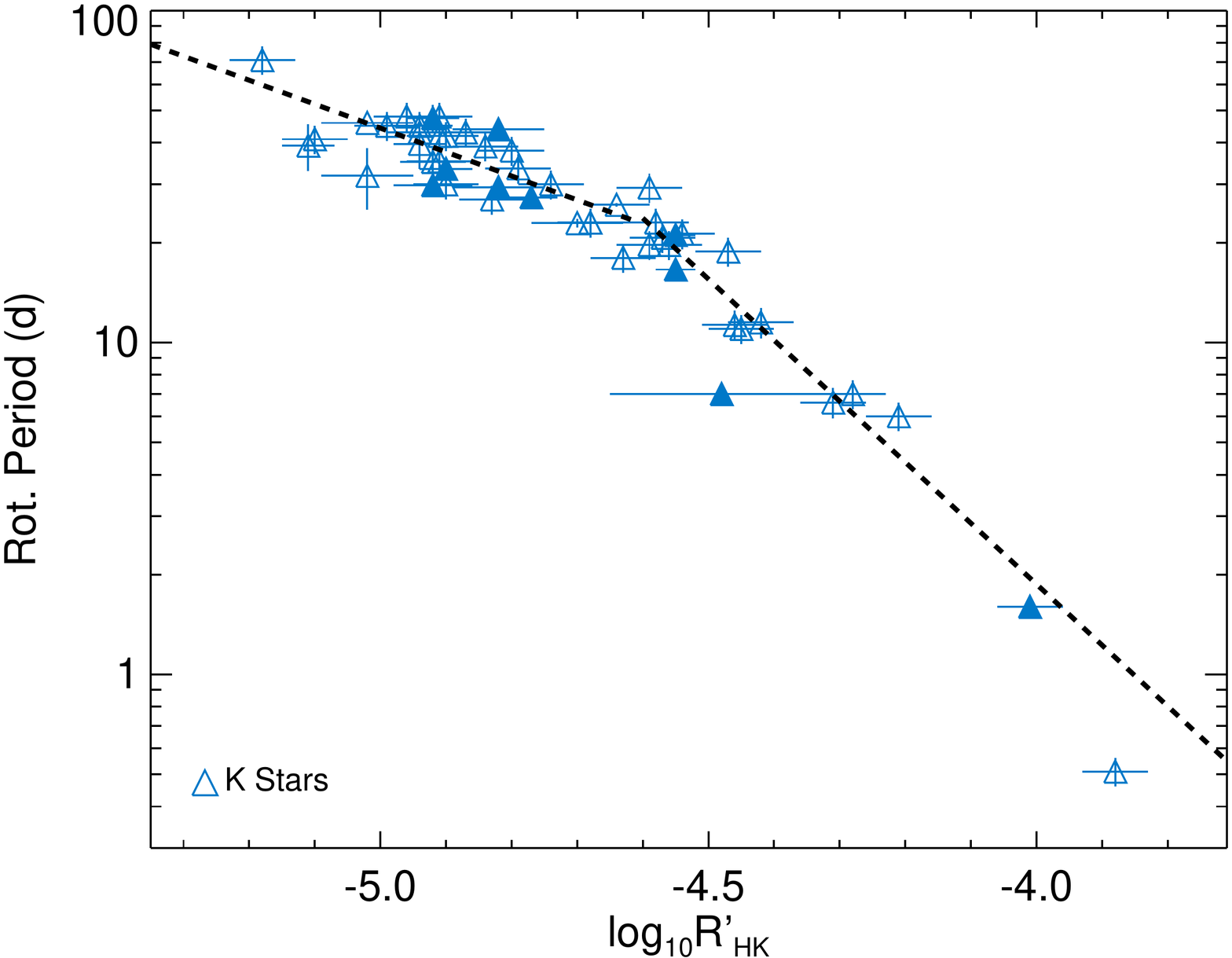}
\includegraphics[width=9.0cm]{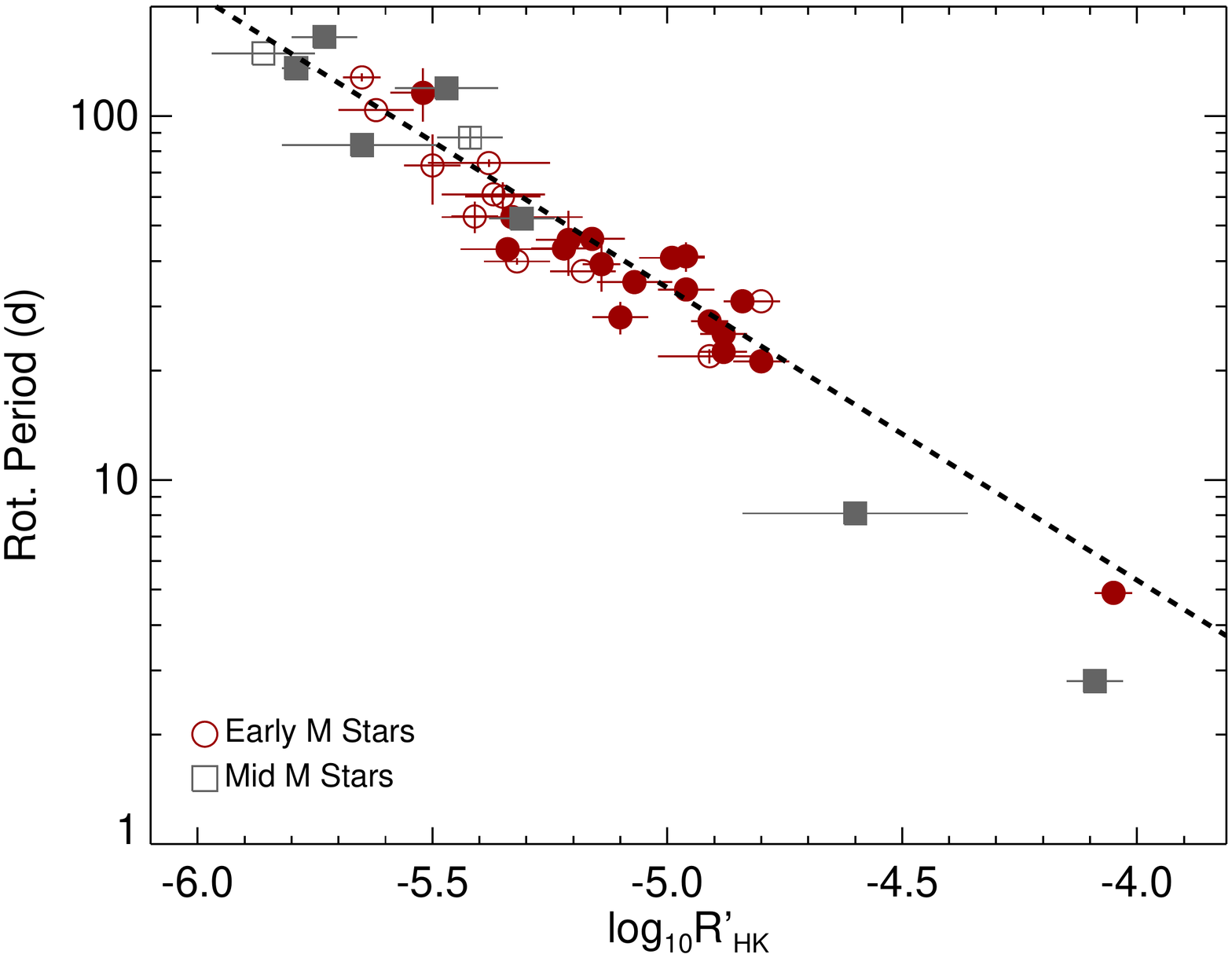}
\caption{Rotation period vs. chromospheric activity level $\log_{10}R'_{HK}$ for each spectral type. Filled symbols show the stars analysed in this work. The dashed line shows the best fit to the data for each individual dataset.}
\label{rhk_period_sp}
\end{figure}

\citet{Masca2015} proposed a rotation-activity relation for late-F- to mid-M-type  main-sequence stars with $\log_{10}R'_{HK}$ $\sim -4.50$ up to $\sim -5.85$. The new measurements presented in Tables~\ref{data_sample} and ~\ref{Periods}, and the data from \citet{Noyes1984}, \citet{Baliunas1995}, and \citet{Lovis2011} serve to extend and better define the relationship. Figure~\ref{rhk_period} shows the new measurements along with those the literature. These measurements are compatible for almost all spectral types and levels of chromospheric activity, as F-type stars are the only clear outlayers. This speaks of a relation that is more complex than what was proposed. Combining all data we can extend the relationship to faster rotators with levels of chromospheric activity up to $\log_{10}R'_{HK}$ $\sim -4$ and we are able to give an independent relationship for each spectral type. Figure~\ref{rhk_period_sp} shows the measurements for every individual spectral type. 

Assuming a relationship  such as
\begin{equation}
\begin{split}
  \log_{10}(P_{rot})= A + B \cdot \log_{10}R'_{HK} 
\end{split}
\label{eq_rhk_period}
,\end{equation}

\begin {table*}
\begin{center}
\caption {Parameters for Eq.~\ref{eq_rhk_period} \label{tab:rhk_period_tab}}
    \begin{tabular}{  l l  l  l l } \hline
Dataset  & N &  A & B & $\sigma$ Period \\ 
        &       &   &  &(\%) \\\hline
G-K-M ($\log_{10}R'_{HK} > -4.5$)               & 37 & --10.118 $\pm$ 0.027     & --4.500 $\pm$ 0.006 & 23 \\
G-K-M ($\log_{10}R'_{HK} \leq -4.5$)            & 94 & --2.425 $\pm$ 0.001      & --0.791 $\pm$ 0.001 &  19 \\ \\
F                                                                               & 25 & -3.609 $\pm$ 0.015 &  --0.946 $\pm$ 0.003 & 39 \\
G ($\log_{10}R'_{HK} > -4.6$)                   & 17     & --11.738 $\pm$ 0.052 &  --2.841 $\pm$ 0.011 & 17 \\
G ($\log_{10}R'_{HK} \leq -4.6$)                        & 27     & 0.138 $\pm$ 0.006 &  --0.261 $\pm$ 0.002 & 20 \\
K ($\log_{10}R'_{HK} > -4.6$)                   & 18 & --7.081 $\pm$ 0.030 &  --1.838 $\pm$ 0.007 & 8 \\
K ($\log_{10}R'_{HK} \leq -4.6$)                        & 32 & --1.962 $\pm$ 0.005 &  --0.722 $\pm$ 0.002 & 16 \\
M                                                                               & 38 & --2.490 $\pm$ 0.002  &  --0.804 $\pm$ 0.001 & 18 \\

 \hline
\label{rhk_period_tab}
\end{tabular}  
\end{center}
\end {table*}

where $P_{rot}$ is in days and  the typical residuals of the fit is smaller than 23 per cent of the measured periods for a given level of activity, for stars from G to mid M, and with residuals smaller than 39 per cent in the case of F-type stars. Table~\ref{rhk_period_tab} shows the coefficients of the best fit for every individual dataset. This relationship provides an estimate of the rotation period of stars with low levels of chromospheric activity.

\begin{figure}
\includegraphics[width=9.0cm]{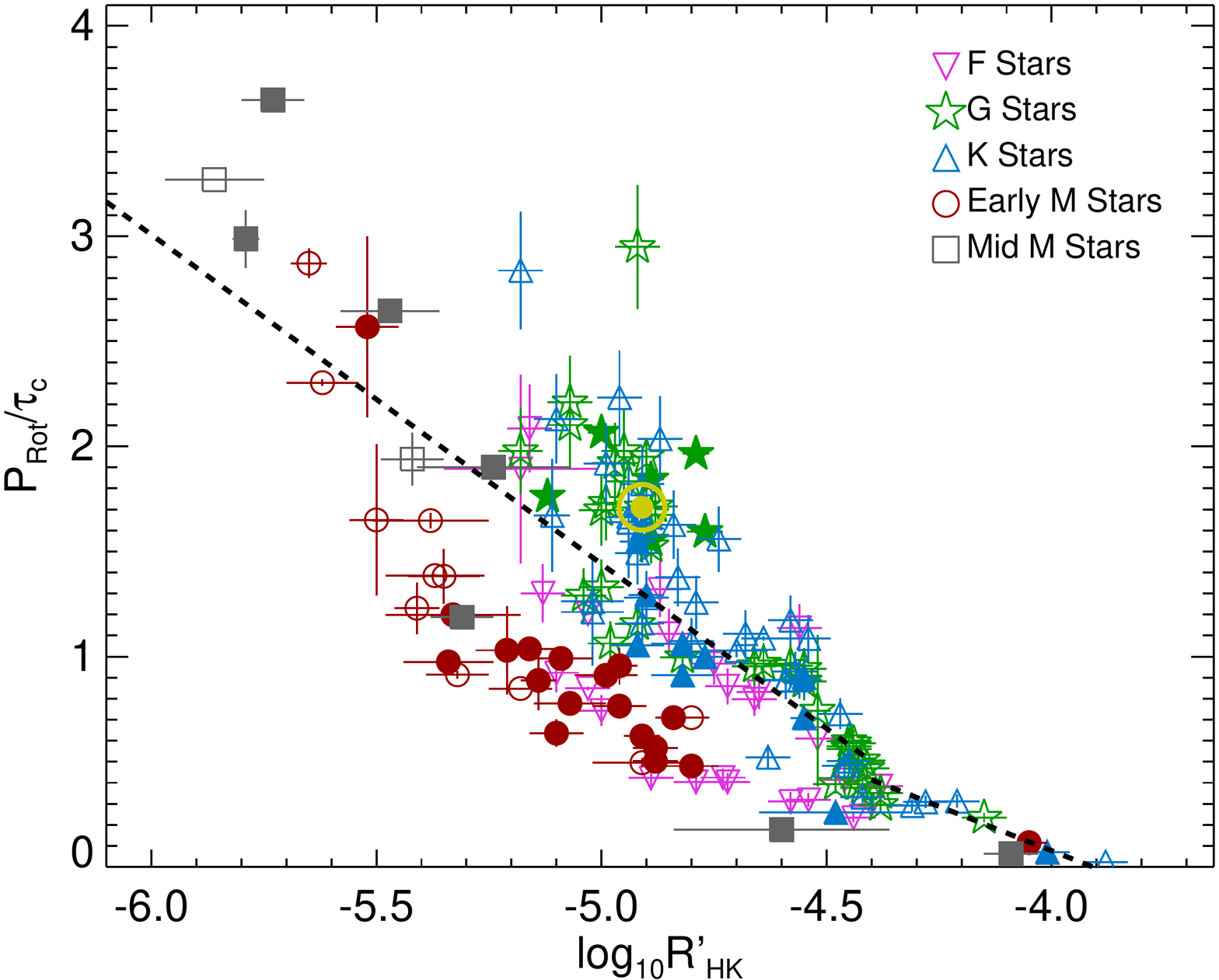}
\caption{Rossby number vs. chromospheric activity level $\log_{10}R'_{HK}$ for each spectral type. Filled symbols show the stars analysed in this work. The dashed line shows the best fit to the data.}
\label{rhk_Rossby}
\end{figure}

The original rotation-activity relationship, which was proposed by \citet{Noyes1984} and updated by \citet{Mamajek2008}, used as their observable the Rossby number -- $Ro = P_{Rot}/\tau_{c}$, i.e. the rotation period divided by the convective turnover,  instead of the rotation period. The use of the convective turnover here raises some problems. The  convective turnover can be determined from theoretical models (e.g. \citet{Gilman1980}, \citet{Gilliland1985}, \citet{Rucinski1986} or \citet{KimDemarque1996}) or empirically (\citet{Noyes1984}, \citet{Stepien1994}, or \citet{Pizzolato2003}). While the values of $\tau_{c}$ are consistent for G-K stars, they diverge badly for M dwarfs. Theoretical models indicate a steep increase of $\tau_{c}$ with decreasing mass, while purely empirical models indicate that $\tau_{c}$ increases with decreasing mass down to 0.8 M$_{\odot}$, but then levels off \citep{Kiraga2007}. The behaviour of $\tau_{c}$  beyond 0.6 M$_{\odot}$ is uncertain. For the measurement of  $\tau_{c}$ we adopt the definition of \citet{Rucinski1986} because this definition produces the tightest correlation, i.e. 

\begin{equation}
\begin{split}
  log (\tau_{c}) = 1.178 - 1.163  x + 0.279  x^2 - 6.14  x^3   ~(x > 0)
  \end{split}
\label{eq_tau}
\end{equation}

\begin{equation}
\begin{split}
  log (\tau_{c}) = 1.178 - 0.941  x + 0.460  x^2  ~(x < 0),
\end{split}
\label{eq_tau}
\end{equation}

where $x = 0.65 - (B-V)$. Figure~\ref{rhk_Rossby} shows the distribution of the calculated Rossby numbers against the $\log_{10}R'_{HK}$. When presenting our results this way we find that, for solar-type stars, the distribution is very similar inside the limits studied in \citet{Mamajek2008}, but moving towards lower levels of activity increases the scatter. On the other hand, F-type and M-type stars do not follow the same exact relationship, hinting that the mechanism might be more complex than originally assumed or that the estimation of the convective turnover could be less reliable the further we get from the Sun, which was already stated by \citet{Noyes1984} for stars with $B-V$ > 1. Unfortunately there is no reliable calibration for the convective turnover in the case of M-dwarf stars. Even with the bigger scatter the global behaviour of the data remains similar with a change of slope when reaching the very active regime ($\log_{10}R'_{HK} \sim -4.4$). 

Analogous to what we did in Eq.~\ref{eq_rhk_period}, we find that the distribution can be described as\begin{equation}
\begin{split}
  Ro =  A + B \cdot \log_{10}R'_{HK} 
\end{split}
\label{eq_rhk_Rossby}
,\end{equation}

where the typical scatter of the residuals is $\sim$ 28 per cent of the measured $Ro$ for the very active region and $\sim$ 19 per cent for the moderately active region. Table~\ref{rhk_Rossby_tab} shows the parameters of equation \ref{eq_rhk_Rossby}. 

\begin {table*}
\begin{center}
\caption {Parameters for Eq.~\ref{eq_rhk_Rossby} \label{tab:rhk_Rossby_tab}}
    \begin{tabular}{  l l  l  l l } \hline
Dataset  & N &  A & B & $\sigma$ Ro \\ 
        &       &   &  &(\%) \\\hline
$\log_{10}R'_{HK} > -4.4$    & 7 & --3.533 $\pm$ 0.796  & --0.912 $\pm$ 0.195 & 28 \\
$\log_{10}R'_{HK} \leq -4.4$ & 150 & --10.431 $\pm$ 0.078       & --2.518 $\pm$ 0.016 &  19 \\
 \hline
\label{rhk_Rossby_tab}
\end{tabular}  
\end{center}
\end {table*}

\subsection{Activity versus photometric cycle amplitude}

A relation between the cycle amplitude in $Ca II_{HK}$ flux variations and the mean activity level of the stars was proposed by \citet{SaarBrandenburg2002}. It was found that stars with a higher mean activity level show also larger cycle amplitudes. We investigate the behaviour of the photometric amplitude of the cycle with the mean activity level as well as with the Rossby number. 

\begin{figure}
\includegraphics[width=9.0cm]{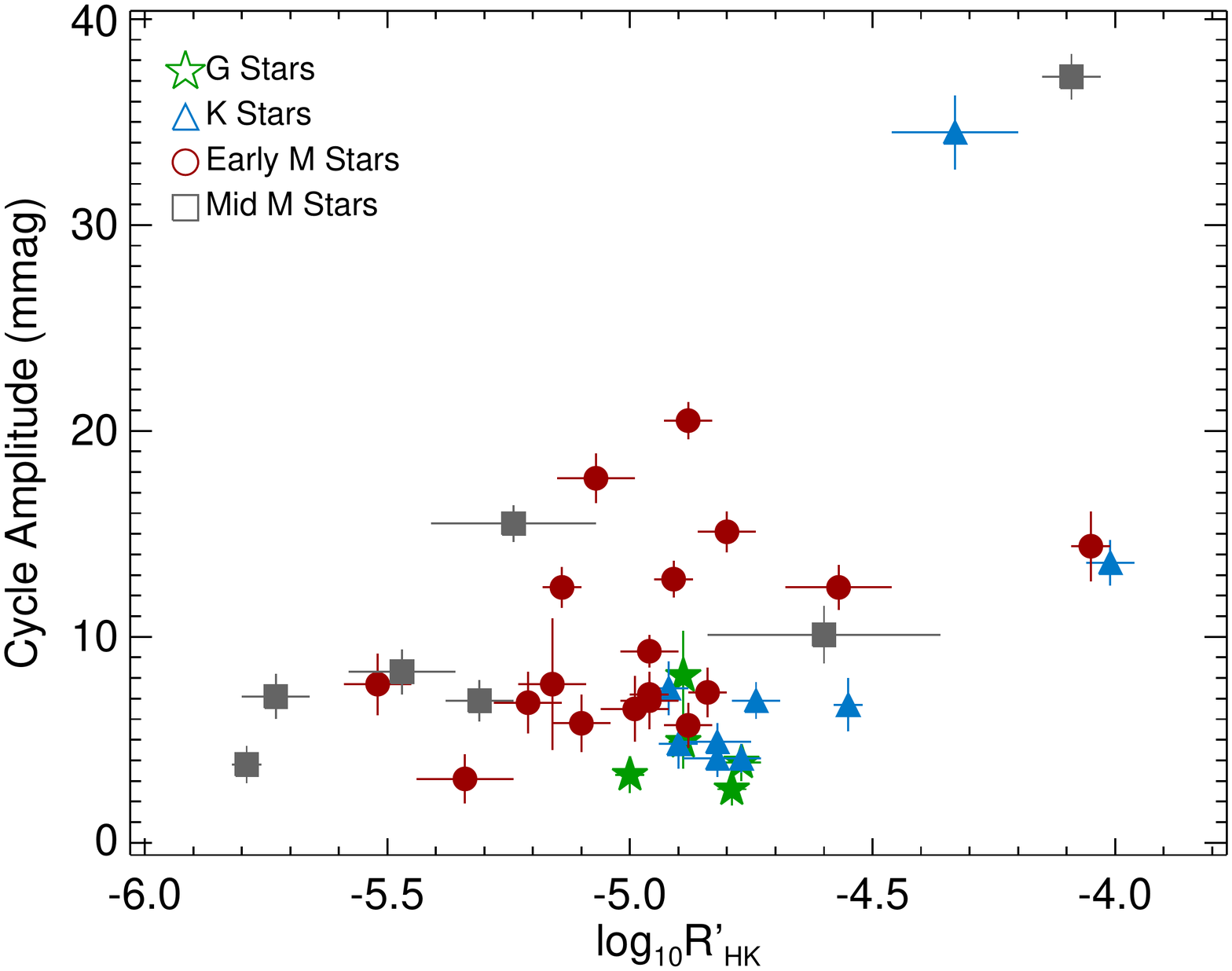}
\caption{Measured cycle photometric amplitude vs. chromospheric activity level $\log_{10}R'_{HK}$.}
\label{rhk_amp_c}
\end{figure}

Comparing the cycle amplitude with the $\log_{10}R'_{HK}$ we are able to see a weak tendency. Even if there is a large scatter, a trend such that the photometric amplitude of the cycle increases towards higher activity stars is found (see Figure~\ref{rhk_amp_c}). This agrees with the \citet{SaarBrandenburg2002} work. 

\begin{figure}
\includegraphics[width=9.0cm]{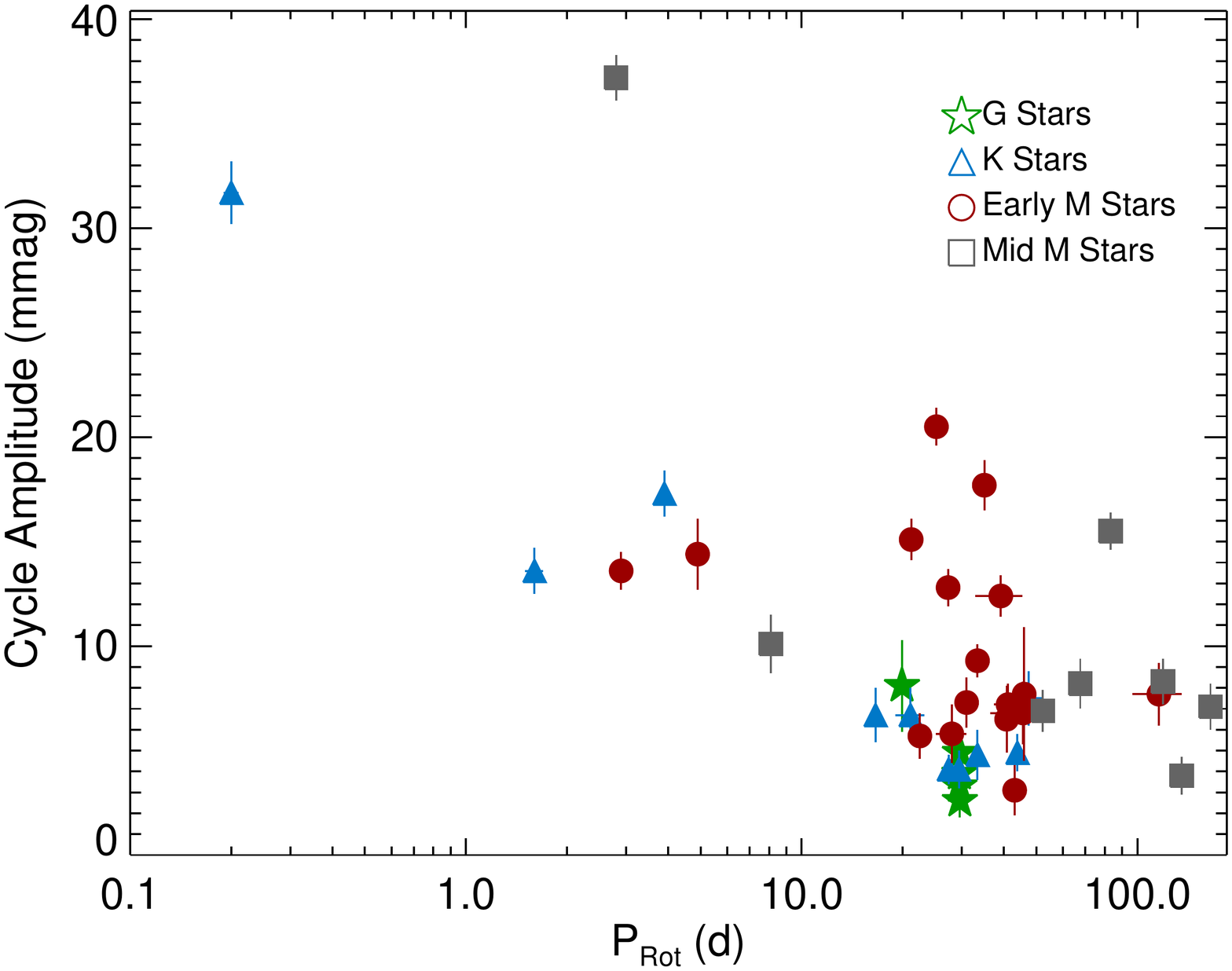}
\caption{Measured cycle photometric amplitude versus the rotation period.}
\label{Rossby}
\end{figure}

When comparing the cycle amplitude with the rotation period we found a more clear correlation. While the scatter is large, is clear that the amplitude decreases longer periods (see Figure~\ref{Rossby}). This is different to what \citet{SaarBrandenburg2002} found when studying the $Ca II$ variations, where cycle amplitude saturation may be happening.  

\subsection{Rotation-cycle relation}

The existence of a relationship between the length of the magnetic cycle and rotation period has been studied for a long time. \citet{Baliunas1996} suggested $P_{cyc}/P_{rot}$ as an observable to study how both quantities relate to each other. It was suggested that the length of the cycle scales as $\sim D^l$, where $l$ is the slope of the relation and $D$ is the dynamo number. Slopes that are different from $\sim 1$ would imply a correlation between the length of the cycle and rotation period. 
 
\begin{figure}
\includegraphics[width=9.0cm]{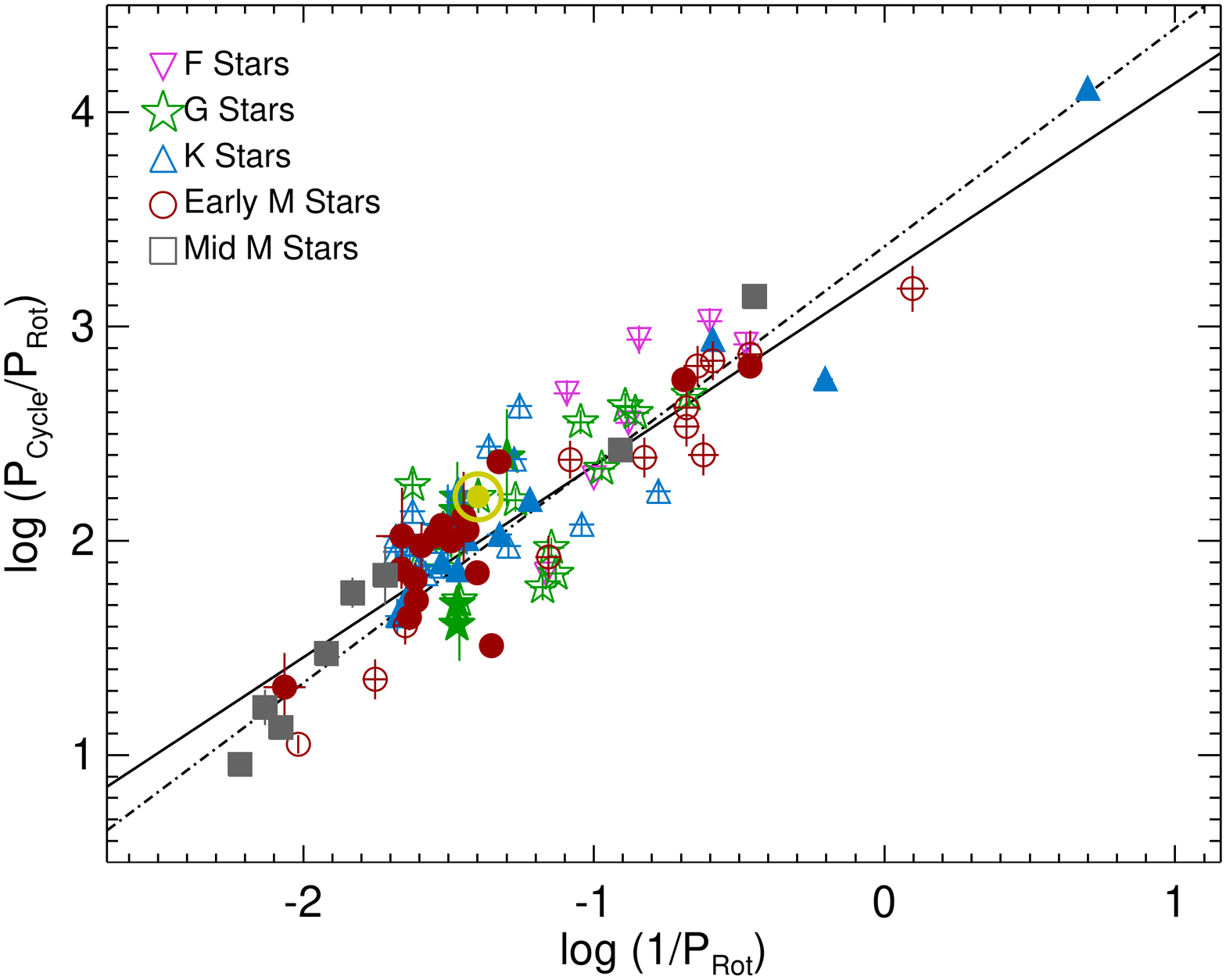}
\caption{$P_{cyc}/P_{rot}$ versus $1/P_{rot}$ in log-log scale. Filled dots represent main-sequence stars while empty circles stand for pre-main-sequence stars. The dashed line shows the fit to the full  dataset. The solid line shows the fit to the main-sequence stars with radiative core.}
\label{cycles}
\end{figure}

Figure~\ref{cycles} shows results on a log-log scale of $P_{cyc}/P_{rot}$ versus $1/P_{rot}$ for all the stars for which we were able to determine both the rotation period and length of the cycle. The slope for our results is $1.01 \pm 0.06$, meaning no correlation between both quantities. However, we cover from early-G to late-M stars, including main-sequence and pre-main-sequence stars, with rotation periods ranging from $\sim 0.2$ to more than $\sim 160$ days.  Previous works that found correlations were based on more homogeneous, and thus suitable samples, to study this relationship. For stars with multiple cycles we choose the global cycle (longer cycle), but the possibility of some of the short period cycles being flip-flop cycles cannot be ruled out.

If we restrict our analysis to main-sequence FGK stars,  we obtain a slope of $0.89\pm 0.05$, meaning that there is a weak correlation between both quantities. The measure is compatible with the result of $0.81 \pm 0.05$ from \citet{Olah2009} and implies a weaker correlation than that found by \citet{Baliunas1996} who found a slope of $0.74$  when sharing a good part of the sample. This supports the idea that there is a common dynamo behaviour for this group of stars, which does not apply  to the M-dwarf stars for which we, as \citet{Savanov2012}, do not find a correlation.  

\begin{figure}
\includegraphics[width=9.0cm]{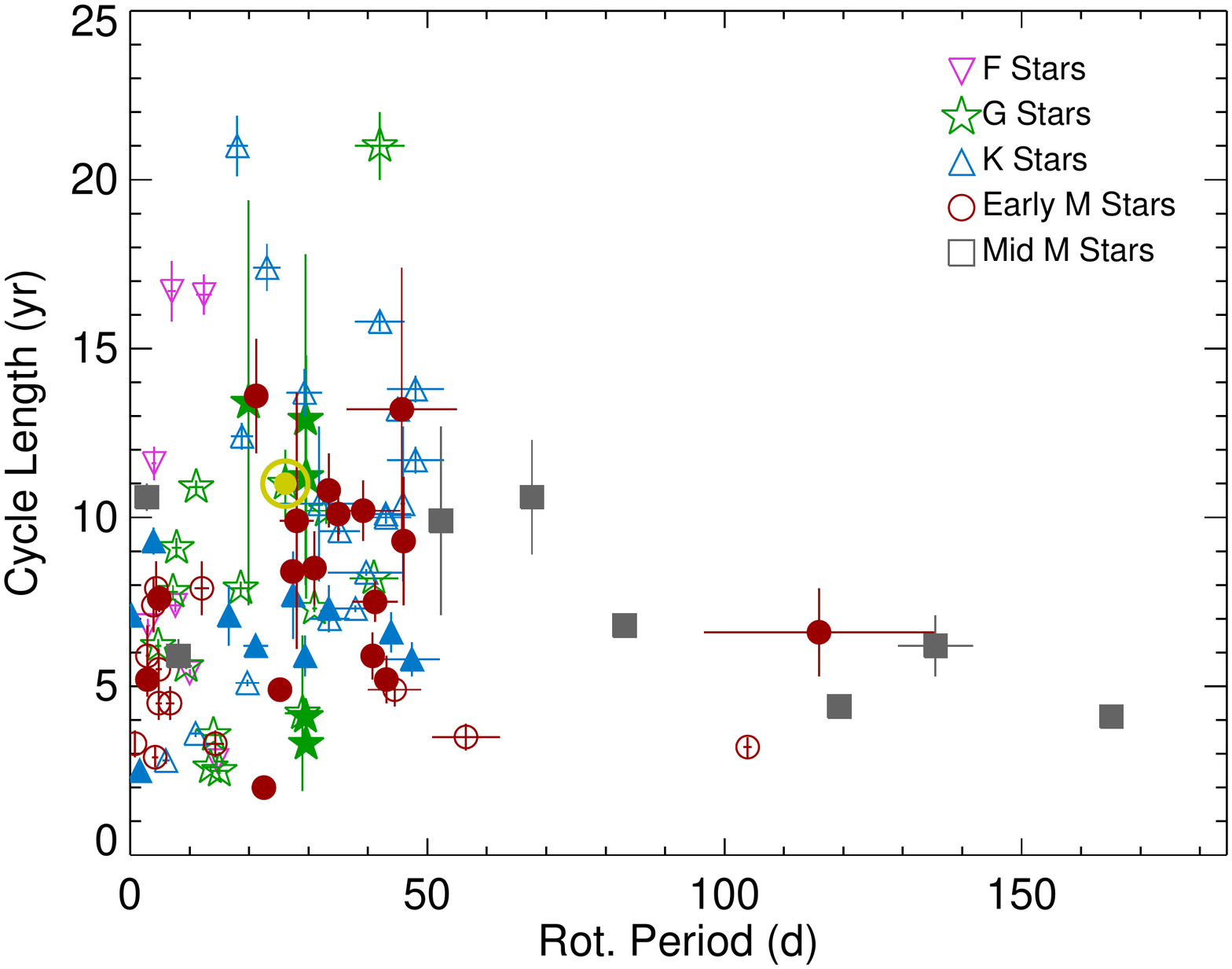}
\caption{Cycle length versus rotation periods. Filled symbols show the stars analysed in this work. }
\label{rot_cycle}
\end{figure}

In a direct comparison of  the length of the cycle with the rotation period, we see an absence of long cycles for extreme slow rotators. Figure~\ref{rot_cycle} shows the distribution of cycle lengths against rotation periods. While the cycle lengths of those stars with rotation periods below the saturation level of $\sim 50$ days distribute approximately uniformly from $\sim 2$ to $\sim 20$ years, those stars rotating slower than $\sim 50$ days show only cycles shorter cycles. At this point, it is unclear whether or not this behaviour is real or is related to a selection or observational bias. The number of stars with both rotation and cycle measured in this region is small, and the time span of the observation is shorter than 10 years. Further investigation is needed to clarify the actual distribution.  

\begin{figure}
\includegraphics[width=9.0cm]{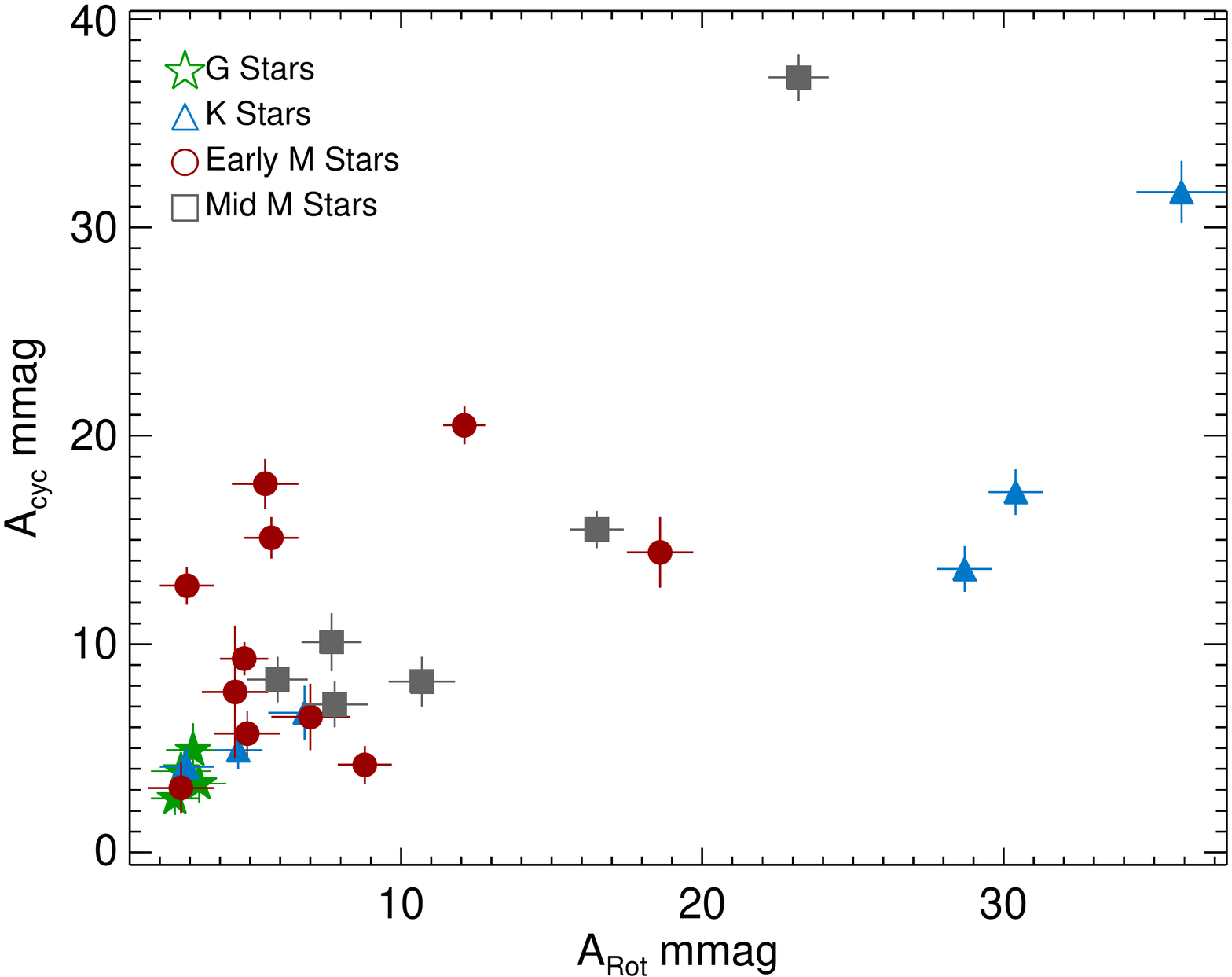}
\caption{Measured photometric amplitude of the cycles vs. the photometric amplitude of the rotational modulation as measured in this work. }
\label{amp_amp}
\end{figure}

Finally, we have also compared the photometric amplitudes induced by the activity cycles and the rotational modulation that seem to show a clear correlation. Both amplitudes increase together with an almost 1:1 relationship, even if with great scatter, and the behaviour seems to be similar for every spectral type (see Fig.~\ref{amp_amp}).  

\section{Summary}

We have analysed the photometric variability in long time series for a sample of solar neighbours observable from the southern hemisphere. We analysed the light curves of a sample of 125 late-A to mid-M stars, containing more than $50\ 000$ observations to measure periodic variabilities induced by the rotation and activity cycles. 

We have been able to identify analogues of the solar cycle for 47 stars of the solar neighbourhood and 36 rotation periods, where 34 of them are M dwarfs.  For 44 stars we have a measurement of the mean activity level ($\log_{10}R'_{HK} $) obtained either from the literate or from our own measurements.

The length of the detected magnetic cycles  goes from over 2.5 years to less than 14 years. No correlation is found between  length of  cycles and the spectral type.   We found some stars that show multiple cycles in which the short-term cycles are  possibly flip-flop cycles.

The length of the measured rotation periods goes from a fraction of a day up to more than 150 days. There seems to be a saturation level at $\sim$ 50 days, only surpassed by M dwarfs, especially fully convective M dwarfs.  

Combining our results with previous measurements we refine the activity-rotation relation proposed in \citet{Masca2015} and extend it up to $\log_{10}R'_{HK} \sim$ -4. The activity-rotation relationship is more complex than previously stated, having a change of slope for activity levels higher than $\log_{10}R'_{HK} \sim$ -4.5 and a completely different behaviour for F-type stars. 

We find evidence for a correlation between the cycle length and the rotation period for the F, G and K main-sequence stars. For these stars the length of the cycle scales as $D^l$ with $l = 0.89$. 

A weak relationships between the activity level and the amplitude of the sun-like cycles can also be found when analysing photometric light curves of G to mid-M dwarfs. A relation is determined between the amplitudes and the rotation period. We find that the faster a star rotates the larger the amplitude of its cycle.   

\begin{acknowledgements}
      This work has been financed by the Spanish Ministry project MINECO AYA2014-56359-P . J.I.G.H. acknowledges financial support from the Spanish MINECO under the 2013 Ram\'on y Cajal program MINECO RYC-2013-14875.
\end{acknowledgements}

%
%
\bibliography{phot_ref}

\begin{thebibliography}{79}
\expandafter\ifx\csname natexlab\endcsname\relax\def\natexlab#1{#1}\fi

\bibitem[{{Baliunas} {et~al.}(1995){Baliunas}, {Donahue}, {Soon}, {Horne},
  {Frazer}, {Woodard-Eklund}, {Bradford}, {Rao}, {Wilson}, {Zhang}, {Bennett},
  {Briggs}, {Carroll}, {Duncan}, {Figueroa}, {Lanning}, {Misch}, {Mueller},
  {Noyes}, {Poppe}, {Porter}, {Robinson}, {Russell}, {Shelton}, {Soyumer},
  {Vaughan}, \& {Whitney}}]{Baliunas1995}
{Baliunas}, S.~L., {Donahue}, R.~A., {Soon}, W.~H., {et~al.} 1995, \apj, 438,
  269

\bibitem[{{Baliunas} {et~al.}(1996){Baliunas}, {Nesme-Ribes}, {Sokoloff}, \&
  {Soon}}]{Baliunas1996}
{Baliunas}, S.~L., {Nesme-Ribes}, E., {Sokoloff}, D., \& {Soon}, W.~H. 1996,
  \apj, 460, 848

\bibitem[{{Baliunas} \& {Vaughan}(1985)}]{Baliunas1985}
{Baliunas}, S.~L. \& {Vaughan}, A.~H. 1985, \araa, 23, 379

\bibitem[{{Berdyugina} \& {J{\"a}rvinen}(2005)}]{BerdyuginaJarvinen2005}
{Berdyugina}, S.~V. \& {J{\"a}rvinen}, S.~P. 2005, Astronomische Nachrichten,
  326, 283

\bibitem[{{Berdyugina} \& {Tuominen}(1998)}]{BerdyuginaTuominen1998}
{Berdyugina}, S.~V. \& {Tuominen}, I. 1998, \aap, 336, L25

\bibitem[{{Berdyugina} \& {Usoskin}(2003)}]{BerdyuginaUsoskin2003}
{Berdyugina}, S.~V. \& {Usoskin}, I.~G. 2003, \aap, 405, 1121

\bibitem[{{Chugainov}(1971)}]{Chugainov1971}
{Chugainov}, P.~F. 1971, Information Bulletin on Variable Stars, 520, 1

\bibitem[{{Chugainov}(1974)}]{Chugainov1974}
{Chugainov}, P.~F. 1974, Izvestiya Ordena Trudovogo Krasnogo Znameni Krymskoj
  Astrofizicheskoj Observatorii, 52, 3

\bibitem[{{Cincunegui} {et~al.}(2007){Cincunegui}, {D{\'{\i}}az}, \&
  {Mauas}}]{Cincunegui2007}
{Cincunegui}, C., {D{\'{\i}}az}, R.~F., \& {Mauas}, P.~J.~D. 2007, \aap, 461,
  1107

\bibitem[{{Cousins} \& {Stoy}(1962)}]{Cousins1962}
{Cousins}, A.~W.~J. \& {Stoy}, R.~H. 1962, Royal Greenwich Observatory
  Bulletins, 64, 103

\bibitem[{{Cumming}(2004)}]{Cumming2004}
{Cumming}, A. 2004, MNRAS, 354, 1165

\bibitem[{{Donahue} {et~al.}(1996){Donahue}, {Saar}, \&
  {Baliunas}}]{Donahue1996}
{Donahue}, R.~A., {Saar}, S.~H., \& {Baliunas}, S.~L. 1996, ApJ, 466, 384

\bibitem[{{Drake} {et~al.}(1989){Drake}, {Simon}, \& {Linsky}}]{Drake1989}
{Drake}, S.~A., {Simon}, T., \& {Linsky}, J.~L. 1989, \apjs, 71, 905

\bibitem[{{Ducati}(2002)}]{Ducati2002}
{Ducati}, J.~R. 2002, VizieR Online Data Catalog, 2237, 0

\bibitem[{{Dumusque} {et~al.}(2012){Dumusque}, {Pepe}, {Lovis},
  {S{\'e}gransan}, {Sahlmann}, {Benz}, {Bouchy}, {Mayor}, {Queloz}, {Santos},
  \& {Udry}}]{Dumusque2012}
{Dumusque}, X., {Pepe}, F., {Lovis}, C., {et~al.} 2012, \nat, 491, 207

\bibitem[{{Dumusque} {et~al.}(2011){Dumusque}, {Santos}, {Udry}, {Lovis}, \&
  {Bonfils}}]{Dumusque2011}
{Dumusque}, X., {Santos}, N.~C., {Udry}, S., {Lovis}, C., \& {Bonfils}, X.
  2011, \aap, 527, A82

\bibitem[{{Fabricius} {et~al.}(2002){Fabricius}, {H{\o}g}, {Makarov}, {Mason},
  {Wycoff}, \& {Urban}}]{Fabricius2002}
{Fabricius}, C., {H{\o}g}, E., {Makarov}, V.~V., {et~al.} 2002, \aap, 384, 180

\bibitem[{{Garc{\'e}s} {et~al.}(2011){Garc{\'e}s}, {Catal{\'a}n}, \&
  {Ribas}}]{Garces2011}
{Garc{\'e}s}, A., {Catal{\'a}n}, S., \& {Ribas}, I. 2011, \aap, 531, A7

\bibitem[{{Gilliland}(1985)}]{Gilliland1985}
{Gilliland}, R.~L. 1985, \apj, 299, 286

\bibitem[{{Gilman}(1980)}]{Gilman1980}
{Gilman}, P.~A. 1980, in Lecture Notes in Physics, Berlin Springer Verlag, Vol.
  114, IAU Colloq. 51: Stellar Turbulence, ed. D.~F. {Gray} \& J.~L. {Linsky},
  19--37

\bibitem[{{Gomes da Silva} {et~al.}(2012){Gomes da Silva}, {Santos}, {Bonfils},
  {Delfosse}, {Forveille}, {Udry}, {Dumusque}, \& {Lovis}}]{GomesdaSilva2012}
{Gomes da Silva}, J., {Santos}, N.~C., {Bonfils}, X., {et~al.} 2012, \aap, 541,
  A9

\bibitem[{{Hall} \& {Henry}(1994)}]{HallHenry1994}
{Hall}, D.~S. \& {Henry}, G.~W. 1994, International Amateur-Professional
  Photoelectric Photometry Communications, 55, 51

\bibitem[{{Hempelmann} {et~al.}(1995){Hempelmann}, {Schmitt}, {Schultz},
  {Ruediger}, \& {Stepien}}]{Hempelmann1995}
{Hempelmann}, A., {Schmitt}, J.~H.~M.~M., {Schultz}, M., {Ruediger}, G., \&
  {Stepien}, K. 1995, \aap, 294, 515

\bibitem[{{H{\o}g} {et~al.}(2000){H{\o}g}, {Fabricius}, {Makarov}, {Urban},
  {Corbin}, {Wycoff}, {Bastian}, {Schwekendiek}, \& {Wicenec}}]{Hog2000}
{H{\o}g}, E., {Fabricius}, C., {Makarov}, V.~V., {et~al.} 2000, A$\&$A, 355,
  L27

\bibitem[{{Horne} \& {Baliunas}(1986)}]{HorneBaliunas1986}
{Horne}, J.~H. \& {Baliunas}, S.~L. 1986, ApJ, 302, 757

\bibitem[{{Irwin} {et~al.}(2011){Irwin}, {Berta}, {Burke}, {Charbonneau},
  {Nutzman}, {West}, \& {Falco}}]{Irwin2011}
{Irwin}, J., {Berta}, Z.~K., {Burke}, C.~J., {et~al.} 2011, \apj, 727, 56

\bibitem[{{Jao} {et~al.}(2014){Jao}, {Henry}, {Subasavage}, {Winters}, {Gies},
  {Riedel}, \& {Ianna}}]{JaoHenry2014}
{Jao}, W.-C., {Henry}, T.~J., {Subasavage}, J.~P., {et~al.} 2014, AJ, 147, 21

\bibitem[{{Jenkins} {et~al.}(2009){Jenkins}, {Ramsey}, {Jones}, {Pavlenko},
  {Gallardo}, {Barnes}, \& {Pinfield}}]{Jenkins2009}
{Jenkins}, J.~S., {Ramsey}, L.~W., {Jones}, H.~R.~A., {et~al.} 2009, ApJ, 704,
  975

\bibitem[{{Jetsu}(1993)}]{Jetsu1993}
{Jetsu}, L. 1993, \aap, 276, 345

\bibitem[{{Kim} \& {Demarque}(1996)}]{KimDemarque1996}
{Kim}, Y.-C. \& {Demarque}, P. 1996, \apj, 457, 340

\bibitem[{{Kiraga}(2012)}]{Kiraga2012}
{Kiraga}, M. 2012, Acta Astron., 62, 67

\bibitem[{{Kiraga} \& {Stepien}(2007)}]{Kiraga2007}
{Kiraga}, M. \& {Stepien}, K. 2007, \actaa, 57, 149

\bibitem[{{Koen} {et~al.}(2010){Koen}, {Kilkenny}, {van Wyk}, \&
  {Marang}}]{Koen2010}
{Koen}, C., {Kilkenny}, D., {van Wyk}, F., \& {Marang}, F. 2010, MNRAS, 403,
  1949

\bibitem[{{Kron}(1947)}]{Kron1947}
{Kron}, G.~E. 1947, \pasp, 59, 261

\bibitem[{{Landolt}(1992)}]{Landolt1992}
{Landolt}, A.~U. 1992, \aj, 104, 340

\bibitem[{{Landolt}(2009)}]{Landolt2009}
{Landolt}, A.~U. 2009, \aj, 137, 4186

\bibitem[{{Lovis} {et~al.}(2011){Lovis}, {Dumusque}, {Santos}, {Bouchy},
  {Mayor}, {Pepe}, {Queloz}, {S{\'e}gransan}, \& {Udry}}]{Lovis2011}
{Lovis}, C., {Dumusque}, X., {Santos}, N.~C., {et~al.} 2011, ArXiv e-prints
  [\eprint[arXiv]{1107.5325}]

\bibitem[{{Mamajek} \& {Hillenbrand}(2008)}]{Mamajek2008}
{Mamajek}, E.~E. \& {Hillenbrand}, L.~A. 2008, ApJ, 687, 1264

\bibitem[{{Markwardt}(2009)}]{Markwardt2009}
{Markwardt}, C.~B. 2009, in Astronomical Society of the Pacific Conference
  Series, Vol. 411, Astronomical Data Analysis Software and Systems XVIII, ed.
  D.~A. {Bohlender}, D.~{Durand}, \& P.~{Dowler}, 251

\bibitem[{{Mayor} {et~al.}(2003){Mayor}, {Pepe}, {Queloz}, {Bouchy},
  {Rupprecht}, {Lo Curto}, {Avila}, {Benz}, {Bertaux}, {Bonfils}, {Dall},
  {Dekker}, {Delabre}, {Eckert}, {Fleury}, {Gilliotte}, {Gojak}, {Guzman},
  {Kohler}, {Lizon}, {Longinotti}, {Lovis}, {Megevand}, {Pasquini}, {Reyes},
  {Sivan}, {Sosnowska}, {Soto}, {Udry}, {van Kesteren}, {Weber}, \&
  {Weilenmann}}]{Mayor2003}
{Mayor}, M., {Pepe}, F., {Queloz}, D., {et~al.} 2003, The Messenger, 114, 20

\bibitem[{{Mekkaden}(1985)}]{Mekkaden1985}
{Mekkaden}, M.~V. 1985, \apss, 117, 381

\bibitem[{{Mermilliod}(1986)}]{Mermilliod1986}
{Mermilliod}, J.-C. 1986, Catalogue of Eggen's UBV data., 0 (1986), 0

\bibitem[{{Middelkoop} {et~al.}(1981){Middelkoop}, {Vaughan}, \&
  {Preston}}]{Middelkoop1981}
{Middelkoop}, F., {Vaughan}, A.~H., \& {Preston}, G.~W. 1981, \aap, 96, 401

\bibitem[{{Montes} {et~al.}(2004){Montes}, {Crespo-Chac{\'o}n}, {G{\'a}lvez},
  {Fern{\'a}ndez-Figueroa}, {L{\'o}pez-Santiago}, {de Castro}, {Cornide}, \&
  {Hern{\'a}n-Obispo}}]{Montes2004}
{Montes}, D., {Crespo-Chac{\'o}n}, I., {G{\'a}lvez}, M.~C., {et~al.} 2004,
  Lecture Notes and Essays in Astrophysics, 1, 119

\bibitem[{{Moss}(2004)}]{Moss2004}
{Moss}, D. 2004, \mnras, 352, L17

\bibitem[{{Noyes} {et~al.}(1984){Noyes}, {Hartmann}, {Baliunas}, {Duncan}, \&
  {Vaughan}}]{Noyes1984}
{Noyes}, R.~W., {Hartmann}, L.~W., {Baliunas}, S.~L., {Duncan}, D.~K., \&
  {Vaughan}, A.~H. 1984, ApJ, 279, 763

\bibitem[{{Ol{\'a}h} {et~al.}(2009){Ol{\'a}h}, {Koll{\'a}th}, {Granzer},
  {Strassmeier}, {Lanza}, {J{\"a}rvinen}, {Korhonen}, {Baliunas}, {Soon},
  {Messina}, \& {Cutispoto}}]{Olah2009}
{Ol{\'a}h}, K., {Koll{\'a}th}, Z., {Granzer}, T., {et~al.} 2009, \aap, 501, 703

\bibitem[{{Olmedo} {et~al.}(2013){Olmedo}, {Ch{\'a}vez}, {Bertone}, \& {De la
  Luz}}]{OlmedoChavez2013}
{Olmedo}, M., {Ch{\'a}vez}, M., {Bertone}, E., \& {De la Luz}, V. 2013, \pasp,
  125, 1436

\bibitem[{{Pallavicini} {et~al.}(1981){Pallavicini}, {Golub}, {Rosner},
  {Vaiana}, {Ayres}, \& {Linsky}}]{Pallavicini1981}
{Pallavicini}, R., {Golub}, L., {Rosner}, R., {et~al.} 1981, \apj, 248, 279

\bibitem[{{Pepe} {et~al.}(2013){Pepe}, {Cristiani}, {Rebolo}, {Santos},
  {Dekker}, {M{\'e}gevand}, {Zerbi}, {Cabral}, {Molaro}, {Di Marcantonio},
  {Abreu}, {Affolter}, {Aliverti}, {Allende Prieto}, {Amate}, {Avila},
  {Baldini}, {Bristow}, {Broeg}, {Cirami}, {Coelho}, {Conconi}, {Coretti},
  {Cupani}, {D'Odorico}, {De Caprio}, {Delabre}, {Dorn}, {Figueira}, {Fragoso},
  {Galeotta}, {Genolet}, {Gomes}, {Gonz{\'a}lez Hern{\'a}ndez}, {Hughes},
  {Iwert}, {Kerber}, {Landoni}, {Lizon}, {Lovis}, {Maire}, {Mannetta},
  {Martins}, {Monteiro}, {Oliveira}, {Poretti}, {Rasilla}, {Riva}, {Santana
  Tschudi}, {Santos}, {Sosnowska}, {Sousa}, {Span{\`o}}, {Tenegi}, {Toso},
  {Vanzella}, {Viel}, \& {Zapatero Osorio}}]{Pepe2013}
{Pepe}, F., {Cristiani}, S., {Rebolo}, R., {et~al.} 2013, The Messenger, 153, 6

\bibitem[{{Pepe} {et~al.}(2011){Pepe}, {Lovis}, {S{\'e}gransan}, {Benz},
  {Bouchy}, {Dumusque}, {Mayor}, {Queloz}, {Santos}, \& {Udry}}]{Pepe2011}
{Pepe}, F., {Lovis}, C., {S{\'e}gransan}, D., {et~al.} 2011, \aap, 534, A58

\bibitem[{{Pizzolato} {et~al.}(2003){Pizzolato}, {Maggio}, {Micela},
  {Sciortino}, \& {Ventura}}]{Pizzolato2003}
{Pizzolato}, N., {Maggio}, A., {Micela}, G., {Sciortino}, S., \& {Ventura}, P.
  2003, \aap, 397, 147

\bibitem[{{Pojmanski}(1997)}]{Pojmanski1997}
{Pojmanski}, G. 1997, \actaa, 47, 467

\bibitem[{{Radick} {et~al.}(1990){Radick}, {Skiff}, \& {Lockwood}}]{Radick1990}
{Radick}, R.~R., {Skiff}, B.~A., \& {Lockwood}, G.~W. 1990, \apj, 353, 524

\bibitem[{{Robertson} {et~al.}(2013){Robertson}, {Endl}, {Cochran}, \&
  {Dodson-Robinson}}]{Robertson2013}
{Robertson}, P., {Endl}, M., {Cochran}, W.~D., \& {Dodson-Robinson}, S.~E.
  2013, \apj, 764, 3

\bibitem[{{Robertson} {et~al.}(2014){Robertson}, {Mahadevan}, {Endl}, \&
  {Roy}}]{Robertson2014}
{Robertson}, P., {Mahadevan}, S., {Endl}, M., \& {Roy}, A. 2014, Science, 345,
  440

\bibitem[{{Rucinski} \& {Vandenberg}(1986)}]{Rucinski1986}
{Rucinski}, S.~M. \& {Vandenberg}, D.~A. 1986, \pasp, 98, 669

\bibitem[{{Saar} \& {Brandenburg}(1999)}]{SaarBrandenburg1999}
{Saar}, S.~H. \& {Brandenburg}, A. 1999, \apj, 524, 295

\bibitem[{{Saar} \& {Brandenburg}(2002)}]{SaarBrandenburg2002}
{Saar}, S.~H. \& {Brandenburg}, A. 2002, Astronomische Nachrichten, 323, 357

\bibitem[{{Saar} \& {Osten}(1997)}]{Saar1997B}
{Saar}, S.~H. \& {Osten}, R.~A. 1997, \mnras, 284, 803

\bibitem[{{Savanov}(2012)}]{Savanov2012}
{Savanov}, I.~S. 2012, Astronomy Reports, 56, 716

\bibitem[{{Simon} \& {Fekel}(1987)}]{SimonFekel1987}
{Simon}, T. \& {Fekel}, Jr., F.~C. 1987, \apj, 316, 434

\bibitem[{{Skumanich}(1972)}]{Skumanich1972}
{Skumanich}, A. 1972, \apj, 171, 565

\bibitem[{{Stauffer}(1984)}]{Stauffer1984}
{Stauffer}, J.~R. 1984, \apj, 280, 189

\bibitem[{{Stepien}(1994)}]{Stepien1994}
{Stepien}, K. 1994, \aap, 292, 191

\bibitem[{{Strassmeier}(2009)}]{Strassmeier2009}
{Strassmeier}, K.~G. 2009, \aapr, 17, 251

\bibitem[{{Strassmeier} {et~al.}(1997){Strassmeier}, {Bartus}, {Cutispoto}, \&
  {Rodono}}]{Strassmeier1997}
{Strassmeier}, K.~G., {Bartus}, J., {Cutispoto}, G., \& {Rodono}, M. 1997,
  \aaps, 125, 11

\bibitem[{{Strassmeier} {et~al.}(1993){Strassmeier}, {Hall}, {Fekel}, \&
  {Scheck}}]{Strassmeier1993}
{Strassmeier}, K.~G., {Hall}, D.~S., {Fekel}, F.~C., \& {Scheck}, M. 1993,
  \aaps, 100, 173

\bibitem[{{Su{\'a}rez Mascare{\~n}o} {et~al.}(2015){Su{\'a}rez Mascare{\~n}o},
  {Rebolo}, {Gonz{\'a}lez Hern{\'a}ndez}, \& {Esposito}}]{Masca2015}
{Su{\'a}rez Mascare{\~n}o}, A., {Rebolo}, R., {Gonz{\'a}lez Hern{\'a}ndez},
  J.~I., \& {Esposito}, M. 2015, \mnras, 452, 2745

\bibitem[{{Torres} {et~al.}(2006){Torres}, {Quast}, {da Silva}, {de La Reza},
  {Melo}, \& {Sterzik}}]{Torres2006}
{Torres}, C.~A.~O., {Quast}, G.~R., {da Silva}, L., {et~al.} 2006, \aap, 460,
  695

\bibitem[{{van Altena} {et~al.}(1995){van Altena}, {Lee}, \&
  {Hoffleit}}]{vanAltena1995}
{van Altena}, W.~F., {Lee}, J.~T., \& {Hoffleit}, E.~D. 1995, {The general
  catalogue of trigonometric [stellar] parallaxes}

\bibitem[{{Vaughan} {et~al.}(1981){Vaughan}, {Preston}, {Baliunas}, {Hartmann},
  {Noyes}, {Middelkoop}, \& {Mihalas}}]{Vaughan1981}
{Vaughan}, A.~H., {Preston}, G.~W., {Baliunas}, S.~L., {et~al.} 1981, \apj,
  250, 276

\bibitem[{{Vida} {et~al.}(2014){Vida}, {Ol{\'a}h}, \& {Szab{\'o}}}]{Vida2014}
{Vida}, K., {Ol{\'a}h}, K., \& {Szab{\'o}}, R. 2014, \mnras, 441, 2744

\bibitem[{{Vilhu}(1984)}]{Vilhu1984}
{Vilhu}, O. 1984, \aap, 133, 117

\bibitem[{{Walter} \& {Bowyer}(1981)}]{WalterBowyer1981}
{Walter}, F.~M. \& {Bowyer}, S. 1981, \apj, 245, 671

\bibitem[{{Ward-Duong} {et~al.}(2015){Ward-Duong}, {Patience}, {De Rosa},
  {Bulger}, {Rajan}, {Goodwin}, {Parker}, {McCarthy}, \&
  {Kulesa}}]{Ward-Duong2015}
{Ward-Duong}, K., {Patience}, J., {De Rosa}, R.~J., {et~al.} 2015, \mnras, 449,
  2618

\bibitem[{{West} {et~al.}(2015){West}, {Weisenburger}, {Irwin},
  {Berta-Thompson}, {Charbonneau}, {Dittmann}, \& {Pineda}}]{West2015}
{West}, A.~A., {Weisenburger}, K.~L., {Irwin}, J., {et~al.} 2015, \apj, 812, 3

\bibitem[{{Wright} {et~al.}(2011){Wright}, {Drake}, {Mamajek}, \&
  {Henry}}]{Wright2011}
{Wright}, N.~J., {Drake}, J.~J., {Mamajek}, E.~E., \& {Henry}, G.~W. 2011,
  \apj, 743, 48

\bibitem[{{Zechmeister} \& {K{\"u}rster}(2009)}]{Zechmeister2009}
{Zechmeister}, M. \& {K{\"u}rster}, M. 2009, A$\&$A, 496, 577

\end{thebibliography}
\label{lastpage}

\end{document}